\documentclass[reprint,amsmath,amssymb,aps]{revtex4-2}
\usepackage{color}
\usepackage{graphicx}
\usepackage{dcolumn}
\usepackage{bm}
\usepackage{CJKutf8}
\usepackage{upgreek}
\usepackage{hyperref}
\begin{document}
	\hypersetup{hidelinks}
	\title{The Inverse-Square Law Force between Vapor-Mediated Droplets}

\author{Zhi Wu Jiang\textsuperscript{1}}
\author{Hang Ding\textsuperscript{1}}
\author{Er Qiang Li\textsuperscript{1}}
\email{eqli@ustc.edu.cn}

\affiliation{\textsuperscript{\rm{1}}Department of Modern Mechanics, 
University of Science and Technology of China, Hefei 230027, China}

\begin{abstract}
{\color{black}In 1687, Sir Issac Newton} published \emph{The Mathematical Principles 
of Natural Philosophy} \cite{newton1987philosophiae} in which 
the law of universal gravitation was derived. 
It is the first inverse-square law discovered in nature, 
combined with Coulomb's law in 1785 \cite{depremier}, 
the two famous inverse-square laws become part of the foundation of physics. 
Why does nature prefer inverse-square laws over the laws of other forms? 
The question is still arousing broad discussion, 
and it is an important topic in physics. 
{\color{black}So far, the origin of inverse-square law is still under 
exploration although from the point of reductionism, 
the law of universal gravitation can be treated as the approximation 
of Einstein’s general relativity under weak gravitation \cite{wald2010general},
and Coulomb's law could be derived from quantum electrodynamics \cite{greiner2008quantum}.} 
Here we discover a new inverse-square law between evaporating droplets 
deposited on a high energy solid substrate. 
For binary droplets, 
we show that the evaporation from a source droplet will create 
a surface tension gradient in the precursor film of a target droplet, 
resulting in a long-range inverse-square law force acting on the target droplet,
and that the inverse proportion decay of the source vapor concentration 
in the space essentially contributes to the inverse-square form of the force. 
Furthermore, 
the inverse-square law force here is shown to hold for all 
experimental parameters tested, 
and other systems such as pure-liquid-droplet system and 
thermocapillary system, 
and it satisfies the superposition principle, 
not only suggesting exciting directions for future droplet research and applications, 
but also benefiting understanding of nature's predilection for inverse-square law. 
\end{abstract}

\date{This manuscript was compiled on \today}

\maketitle

\begin{figure*}[t]
\centering
\includegraphics[width=\linewidth]{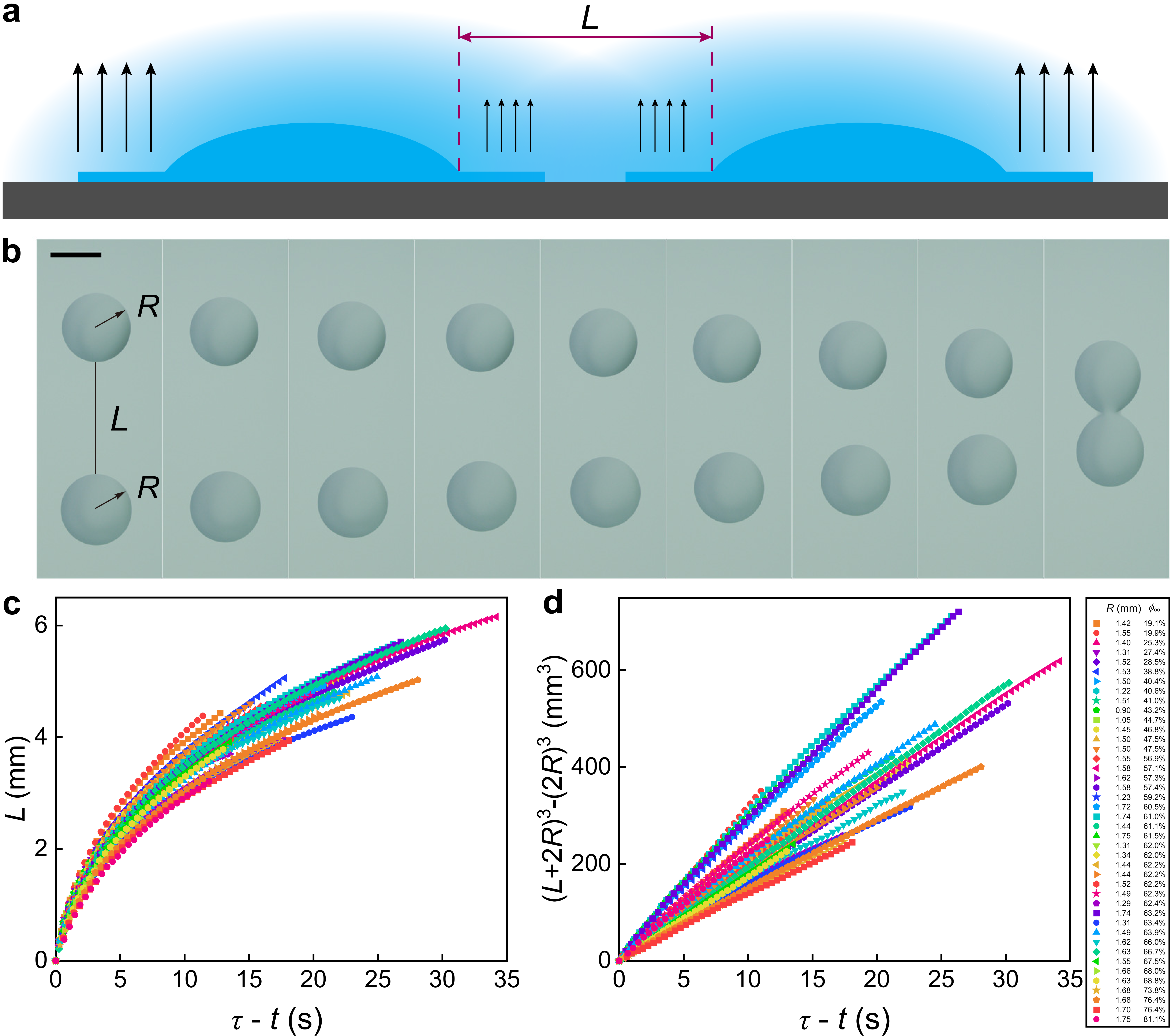}
\caption{{\color{black}Attraction between two binary droplets.} 
(a) A copy sketch to fig. 3a in \cite{cira2015vapour} showing the 
vapor gradients (blue shading) and evaporation (upward arrows) from 
two sessile binary droplets.  
(b) Top view showing attraction of two 10 wt\% PG sessile droplets 
of the same size on a hydrophilic silicon substrate. 
The spreading radius of droplets is $R = 1.37$ mm, 
ambient relative humidity $\phi_\infty$ is 51.3\%,
room temperature is 20$^{\circ}$C, 
the time of droplet contact is $\tau=22$ s. 
Sequence of images spaced by 2.75 s. 
Scale bar is 2 mm. 
(c) Evolution of the distance $L$ between two droplets 
with time $\tau-t$ for various $R$ and $\phi_\infty$. 
(d) All data in (c) fall into straight lines implying a linear 
relationship of $(L + 2R)^3 - (2R)^3 \sim \tau - t$. 
Values of $R$ and $\phi_\infty$ are listed in the legend.}\label{display_all}
\end{figure*}

The inverse-square law states that a physical quantity is inversely 
proportional to the square of the distance {\it r}, 
from the source of that physical quantity. 
Its concise expression greatly advances our understanding of 
fundamental physical principles. 
In fluid mechanics field, 
an analogue of the Newton law of gravity or Coulomb's law was first 
proposed to describe the lateral capillary interaction between 
particles bound to interfaces \cite{paunov1993lateral}. 
Such an interaction force generally originates from the overlap of 
perturbations at interfaces, 
which may be attributed to gravity of floating particles (also known 
as the Cheerios effect \cite{vella2005cheerios}),
wetting of particles \cite{kralchevsky2000capillary}, 
or elastic deformation of soft substrates by droplets (the inverted 
Cheerios effect \cite{karpitschka2016liquid}). 
Paunov {\it et al.} \cite{paunov1993lateral} showed that the lateral 
capillary force obeyed a power law resembling the two-dimensional 
Coulomb's law in certain range of interparticle distances, 
and defined the so-called capillary charge to mimic electric charges. 

This analogue makes us wonder what will happen between two 
evaporating sessile droplets, 
as their evaporating fields overlap like the above mentioned 
perturbations at interfaces do. 
Evaporating sessile droplets abound in nature and technology, 
forming an important ramification of two-phase flows \cite{deegan1997capillary}.
Understanding their behavior can be rewarding for 
advancing interfacial mechanics \cite{de2013capillarity, lohse2020physicochemical} 
and industrial applications \cite{yunker2011suppression, li2019gravitational, li2020evaporating}.
For droplets consisting of miscible liquids, 
they attract if the more volatile component has higher surface 
tension \cite{sadafi2019vapor}. 
As shown in Fig. \ref{display_all}a (a copy sketch to fig. 3a from 
Cira, Benusiglio \& Prakash \cite{cira2015vapour}), 
higher vapor concentration between the droplets leads to less evaporation 
of at proximal sides (e.g. right-side for the left droplet) than distal 
sides of the two droplets. 
Consequently, more volatile component liquid with higher surface tension 
will persist at proximal sides, 
creating a net surface tension force pulling two droplets toward each other. 
Such a movement triggered by surface tension gradients is also referred to 
as Marangoni effect \cite{Daniel2001}. 
For droplets made of pure liquids, 
they could also move, 
as similar droplet arrangement in Fig. \ref{display_all}a will also cause 
nonuniform vapor density \cite{man2017vapor, wen2019vapor}, 
which in turn leads to subsequent variations of the evaporation-induced 
contact angles or the thermocapillarity effect \cite{sadafi2019vapor}. 
Nevertheless, compositional surface tension gradients are usually orders 
of magnitude larger than the thermal ones \cite{hu2005analysis, karpitschka2017marangoni}.

In the above studies, different integral forms rather than analytical 
solutions were used to calculate the net surface tension force 
or the driving force \cite{cira2015vapour, malinowski2020nonmonotonic}, 
which qualitatively matched with experimental data. 
In this study, 
we show for the first time that the interaction force between two 
evaporating sessile droplets is an inverse-square law force and 
validate our theory with high precision experimental data. 

\begin{figure*}[htbp]
\centering
\includegraphics[width=0.89\linewidth]{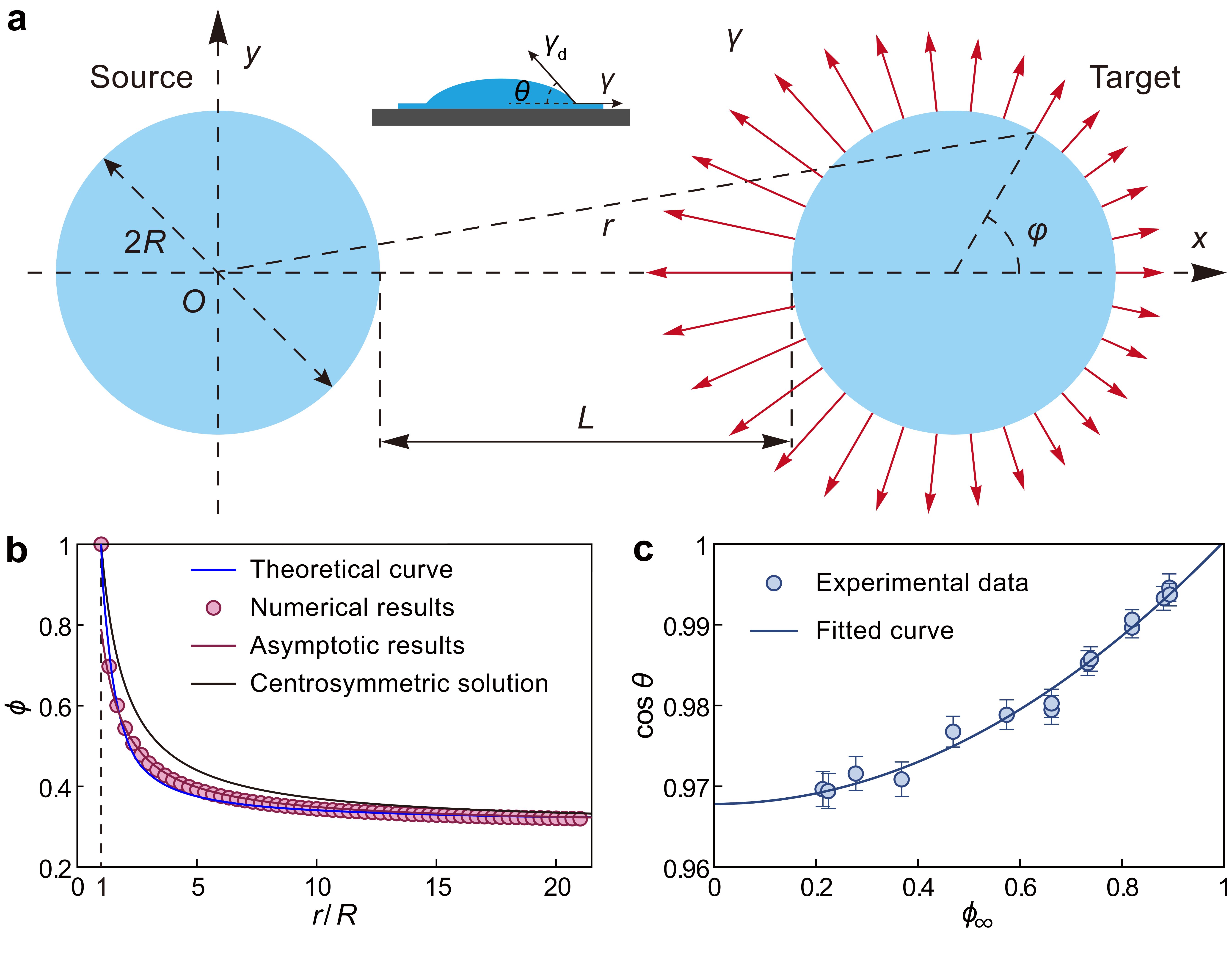}
\vspace*{-0.3in}
\caption{Dependence of contact angle on evaporation from a ``Source'' droplet.
(a) Schematic showing the change in surface tension $\gamma$ along the 
periphery of the ``target'' droplet, 
as a result of evaporation from the ``source'' droplet. 
The inset: $\gamma_d$ the surface tension of the bulk droplet liquid, 
$\gamma$ the surface tension in the precursor liquid film, 
and $\theta$ the contact angle, respectively.
(b) {\color{black}The local relative humidity $\phi$ along the solid substrate. 
$r/R$ is the dimensionless distance. 
Blue curve: theoretical results of $\phi$ (Eq. S10 
in Supplementary Information), 
for the evaporation of a thin film with $\theta$ = 0. 
Dots: numerically solved results of $\phi$ for the evaporation from a sessile droplet. 
Parameters used in the simulation: 
the ambient humidity in the far field $\phi_{\infty} = 30.0\%$, 
$\theta = 10^{\circ}$. 
Red curve: $\phi(r)=\lambda(1-\phi_\infty)R/r+\phi_\infty$, 
an asymptotic solution of Eq. S10 with $\lambda$ = 0.7. 
Black curve: centrosymmetric solution of $\phi(r)$ for the evaporation 
from a droplet floating freely in the air.} 
Liquid is 10 wt\% PG aqueous solution. 
(c) Contact angle as a function of $\phi_{\infty}$ 
for a 10 wt\% PG droplet on a 
hydrophilic silicon surface.}\label{humidity_all} 
\end{figure*}

In our experiments, 
two droplets made of the same propylene glycol (PG) aqueous 
solution were produced from a minuscule capillary tube and 
were dropped onto a pristine oxygen plasma cleaned silicon wafer, 
and the interaction between them was acquired with a high resolution 
digital camera (8K RED DSMC2 Camera, RED Digital Cinema, LLC.). 
To control the surroundings humidity and temperature, 
droplet experiments were carried in an environmental chamber 
(5503-1-11-100-1741 Mini Glove Chamber, Electro-Tech Systems, Inc.) 
which provided transparent walls for good visibility.
A sketch of experimental setup and properties of liquids used in 
the study can be found in Supplementary Information section {\color{black}I}. 
Parameter tested ranges from {\color{black}0.07 - 0.80 $\upmu$L, 
19.9\% - 81.1\%, 
for droplet volume $\it \Omega$ and relative humidity (RH) $\phi_\infty$ 
in the environmental chamber, respectively}. 
When a droplet consisting of miscible liquids is located on a superhydrophilic substrate, 
it has shown that the interplay among Marangoni flow, capillary flow, 
diffusive transport, 
and evaporation will lead to a droplet profile with an apparent nonzero contact angle, 
even though each of the liquid components fully wets the substrate \cite{cira2015vapour, karpitschka2017marangoni, lohse2020physicochemical}. 
This phenomenon is also observed in our experiments 
for 10 wt\% PG droplets (Fig. \ref{humidity_all}c). 
Repeatability test (Supplementary Information Fig. S1) 
shows good reliability of experimental measures. 
Figure \ref{display_all}b shows the attraction and later coalescence 
of two 10 wt\% PG droplets with a distance $L$ apart, 
the mean spreading radius $R$ = 1.37 mm and environmental relative 
humidity $\phi_\infty$ = 51.3 \%. 
Figure \ref{display_all}c shows the change of $L$ verse time $\tau - t$, 
for various trials with $R \in [0.8, 1.8]$ mm and 
$\phi_\infty \in [19.9\%, 81.1\%]$, where $\tau$ is the moment of 
first contact of two droplets. 
Previously experimental data was also presented in such a way \cite{cira2015vapour}, 
which we believe is not conducive to further mechanism exploration. 
Instead, after a transformation, 
Fig. \ref{display_all}d shows a very interesting result that all curves 
transform into straight lines through the origin. 
Despite of the seemingly meaningless arrangement of $(L + 2R)^3 - (2R)^3$, 
in the following we will show the concise and crucial physics implied by 
\begin{equation}
    (L + 2R)^3 - (2R)^3 = k (\tau - t).
    \label{cube_formular}
\end{equation}
The differential form of Eq. \ref{cube_formular} reads 
{\color{black}$-3 dL/dt = k/(L+2R)^2$}, 
or 
\begin{equation}
    {\color{black}-6 U = k/(L+2R)^2} 
    \label{diffstart_formular}
\end{equation}
with the kinematic relationship of $2 U = dL/dt$, 
where $U$ is the velocity of the moving droplet. 
For a Stokes flow considering here \cite{cira2015vapour}, 
the drag force $F_d=3\pi\mu RU l_n/\theta\propto U$ during droplet 
motion is mainly originated from the region near 
the contact line \cite{cira2015vapour,brochard1989motions}, 
and is balanced by the driving force $F$. 
Here $\mu$ denotes the dynamic viscosity of the liquid, 
$\theta$ is the contact angle, 
$l_n$ is the logarithmic cut-off coefficient \cite{de2013capillarity}. 
Consequently, 
the right-hand term in Eq. \ref{diffstart_formular} can be comprehended 
as the driving force with an inverse square law form, 
as $L + 2R$ clearly represents the distance between centers of two droplets! 

The deduction here is quite similar to the method used by Sir Issac Newton 
when he derived the law of universal gravitation from Kepler's three laws 
and the Newton's law of motion proposed by himself. 
Nevertheless, 
the law of universal gravitation is derived from pre-existing indisputably laws, 
here the origin of the inference e.g. Eq. \ref{cube_formular}, 
is just a new empirical formula. 
Therefore, in the following we will thoroughly discuss the new inverse-square 
force from various aspects, including its ``propagator'', precise form, 
and universal applicability. 

\begin{figure*}
\centering
\includegraphics[width=\linewidth]{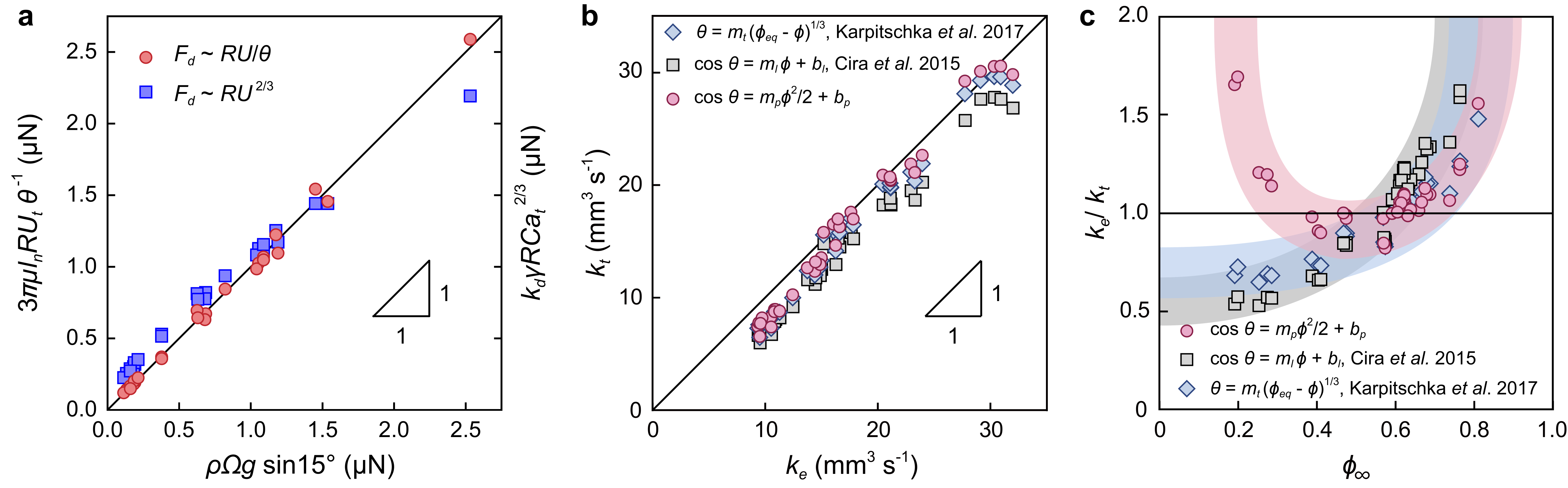}
\caption{{\color{black} Investigation on two models of drag force and 
three models of $\theta = f(\phi)$.} 
(a) Comparison between drag force and gravity force component 
along the substrate. 
The straight line has a slope of 1. 
(b) Comparison between the prefactors $k_t$ calculated theoretically 
from Eq. \ref{kt}, 
and $k_e$ measured experimentally from Fig. \ref{display_all}d, 
for the three different models of contact angle. 
Data are selected for $\phi_{\infty} \sim$ 60\%. 
(c) The ratio of $k_e/k_t$ versus environment humidity $\phi_\infty \in [19.9\%, 81.1\%]$. 
Fitting coefficients are $m_p=0.065$ and $b_p=0.97$, $m_l = 0.036$ and 
$b_l = 0.96$, $m_t = 0.28$ and $\phi_{eq} = 0.95$, 
for the three different models of contact angle, respectively.} 
\label{drag}
\end{figure*}

The origin of the driving force $F$ has been revealed \cite{cira2015vapour} 
and described above. 
{\color{black}Here we propose that evaporated vapour molecules 
could be comprehended as ``propagator'' for vapor-induced driving forces}. 
For PG-water mixture, 
vapor pressure of PG is negligible at room temperature and 
vapor mainly originates from water evaporation. 
Therefore, the local vapor concentration can be replaced by 
relative humidity $\phi(r)$, or $\phi$ for simplicity, 
where $r$ denotes the distance from center of the ``source'' 
droplet (Fig. \ref{humidity_all}a). 
It is shown that $\phi$ satisfies the Laplace equation \cite{hu2002evaporation}, 
$\nabla^2\phi=0$, which has a centrosymmetric solution of 
$\phi(r) = (1 - \phi_\infty) R / r + \phi_\infty$ (black curve 
in Fig. \ref{humidity_all}b), 
for a droplet floating freely in the air, 
where $\phi_\infty$ denotes the ambient humidity far from droplets. 
Here for a spherical segment droplet with a finite contact angle $\theta$,
the Laplace equation was solved numerically with finite element 
method software FreeFem++. 
Despite of the fact that $\phi$ also changes with $\theta$, 
it converges to a single curve when $\theta$ $\leq$ 30$^{\circ}$ (Fig. S3). 
Meanwhile, for a liquid film with $\theta$ = 0$^{\circ}$, 
we are able to derive an analytic solution of $\phi$ 
(blue curve in Fig. \ref{humidity_all}b, 
see detailed derivation in Supplementary Information section II). 
Therefore, we can use Eq. S10 to determine $\phi$ for 
droplets with a contact angle $\theta$ $\leq$ 30$^{\circ}$. 
For simplicity, 
$\phi(r)=\lambda(1-\phi_\infty)R/r+\phi_\infty$, 
an asymptotic solution to Eq. S10 with $\lambda$ = 0.7, 
will be used in later analysis where $\theta \in [5^{\circ}, 15^{\circ}]$. 
The distribution of $\phi(r)$ produces a gradient in $\gamma$, 
the surface tension of the precursor liquid, 
and subsequently the driving force for droplet motion \cite{cira2015vapour}. 
Figure \ref{humidity_all}a schematically shows the change in $\gamma$ 
along the periphery of ``target'' droplet. 

To calculate the force $F$, 
the dependence of $\gamma$ on $\phi(r)$ needs to be pinned down. 
This can be challenging for direct measurement. 
In practice we choose to measure the change of contact angle $\theta$ 
with $\phi_\infty$ as $\gamma$ is directly connected with $\theta$ by 
$\gamma = \gamma_d \cos\theta$, 
where $\gamma_d$ represents the surface tension of the bulk droplet 
liquid (inset figure in Fig. \ref{humidity_all}a). 
The best fit to our experimental data in Fig. \ref{humidity_all}c 
provides an empirical formula $\cos\theta = f(\phi) = m_p\phi^2/2 + b_p$, 
with fitting parameters $m_p$ = 0.065 and $b_p$ = 0.97. 
Different power-laws have also been proposed \cite{cira2015vapour,karpitschka2017marangoni}, 
nevertheless, 
{\color{black}we believe the choice of exponents may affect the concrete form of $F$, 
but will not change its inverse-square nature which is strongly 
implied by Eq. \ref{cube_formular}}. 

Now it is time to calculate $F$. 
The surface tension gradient $\it \Gamma$ reads 
\begin{equation}
\mathit{\Gamma} = \frac{d\gamma}{dr} = \frac{d\gamma}{d\phi}\frac{d\phi}{dr} 
= - \gamma_d m_p\phi_\infty\frac{\lambda(1-\phi_\infty)R}{r^2}.
\label{force_formular_1}
\end{equation}
For droplets apart far away from each other ($r \gg R$), 
assuming $\it \Gamma$ is a constant and taking the value at center of ``target'' droplet, 
with Gauss's flux theorem, 
we get the force along $x$-axis 
\begin{equation}
F = \pi R^2\mathit{\Gamma} = 
- \frac{\gamma_d m_p\phi_\infty\lambda\pi R^{3} 
(1 - \phi_\infty)}{(L + 2R)^{2}}.
\label{force_formular_2}
\end{equation}
It is an inverse-square force!  
Besides above approximate calculation, 
a rigid derivation of $F$ can be found in Supplementary Information section IV, 
in which the exact same form of $F$ is obtained with third order accuracy. 

To verify Eq. \ref{force_formular_2}, 
a direct way is to compare $F$ with the drag force $F_d=3\pi\mu RU l_n/\theta$. 
Before that, 
we first investigate the veracity of $F_d$ and pin down its cut-off coefficient $l_n$. 
Herein droplets were released onto and slided down along a clean silicon wafer 
tilted 15$^{\circ}$ from horizontal line. 
Droplet volume $\it \Omega$ was measured by a Phantom v2512 CMOS high-speed 
video camera from side view. 
Droplet slide motion along silicon wafer was acquired with a RED DSMC2 Camera 
perpendicular to substrate, 
to determine droplet terminal velocity $U_t$ (Fig. S6 in Supplementary Information). 
The gravity force component along wafer 
$F_g = \rho \it{\Omega} g \sin$15$^\circ$, 
where $g$ is the gravity acceleration, 
should be balanced with the drag $F_d = 3\pi\mu R U_t l_n/\theta$ 
once droplet reaches its terminal velocity. 
Figure \ref{drag}a shows an excellent agreement between calculated $F_g$ and $F_d$ (dots in red). 
Cut-off coefficient $l_n$ is determined to be 17.8, 
close to 15 reported for liquid spreading on dry surfaces \cite{hoffman1975study}. 
Figure \ref{drag}a also draws drag force calculated from 
$F_d \sim \gamma R C a^{2 / 3}$ (dots in blue), 
another widely used model based on the Cox-Voinov theory which considers 
apparent contact angle change during motion \cite{voinov1976hydrodynamics, cox1986dynamics, de2013capillarity, reyssat2014drops} (see more details in Supplementary Information section V). 
It is shown that $3\pi\mu R U_t l_n/\theta$ agrees better with experimental data, 
implying amendment may be needed for applying the Cox-Voinov 
theory to scenarios where Marangoni contraction 
is involved in \cite{karpitschka2017marangoni, shiri2021thermal}. 
Giving the model of the drag force, 
the influence of droplet deformation on droplet motion as $\tau - t$ 
approaches zero is investigated and found to be negligible (Fig. S7 
and Supplementary Information section VI). 

\begin{figure*}[tbp]
\centering
\includegraphics[width=0.88\linewidth]{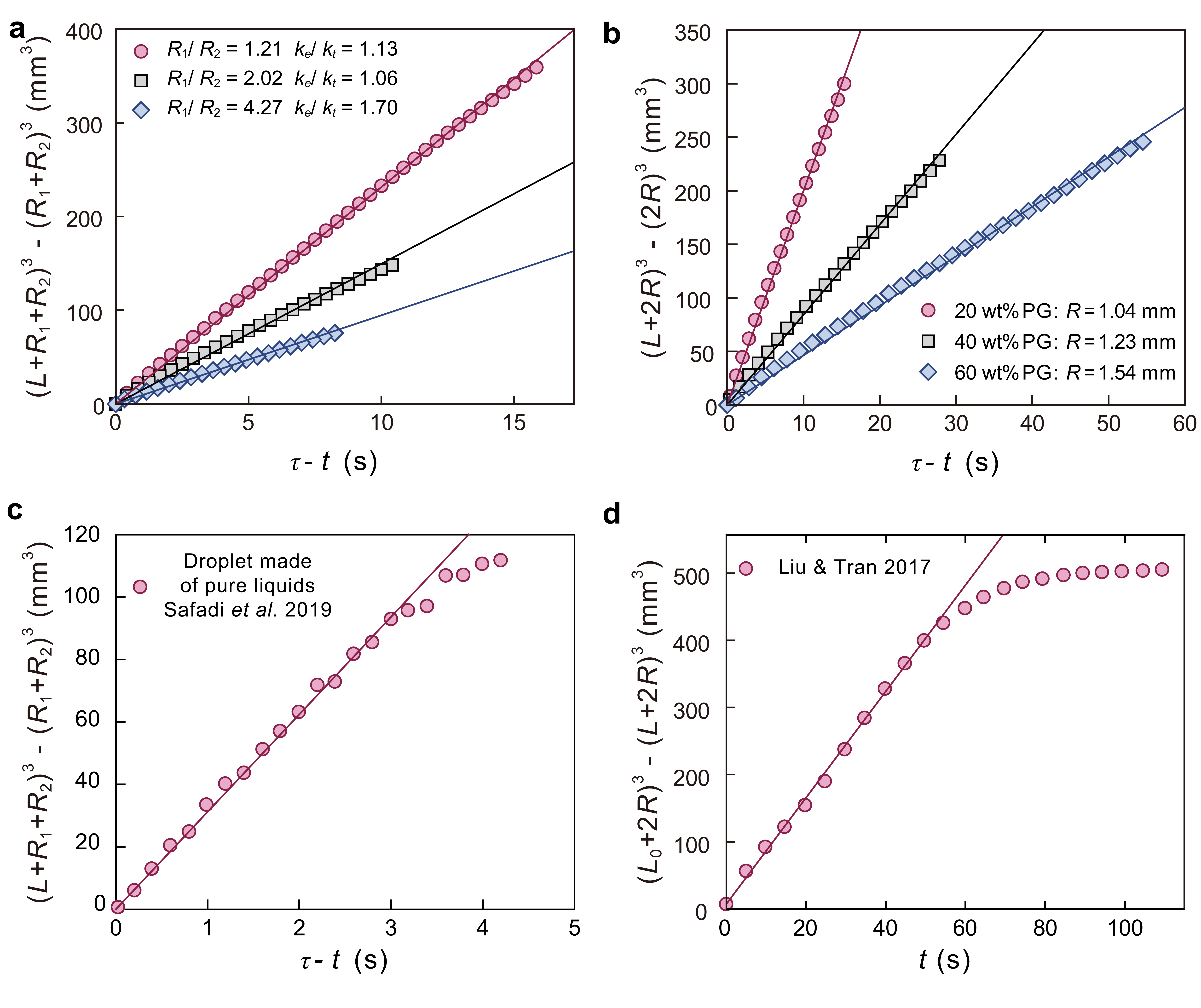}
\vspace*{-0.1in}
\caption{{\color{black}Universal applicability of the inverse-square law force 
between vapor-mediated droplets.} 
Data fall into straight lines implying an inverse-square force between 
(a) 10 wt\% PG droplets of different sizes, 
(b) PG aqueous droplets of different concentrations, 
and (c) droplets made of pure liquid (experimental data extracted from Sadafi {\it et al.} \cite{sadafi2019vapor}). 
(d) Attraction of two 80 wt\% IPA aqueous droplets floating on a silicone oil pool. 
Data in early stage fall into a straight line, 
implying an inverse-square force. 
Deviation occurs when thermally affected zones of two
droplets overlap, which brings other force into play and interferes droplet motion. 
Data are extracted from Liu \& Tran \cite{liu2018vapor}.}\label{extension_mod}  
\end{figure*}

\begin{figure*}[htbp]
\centering
\includegraphics[width=0.8\linewidth]{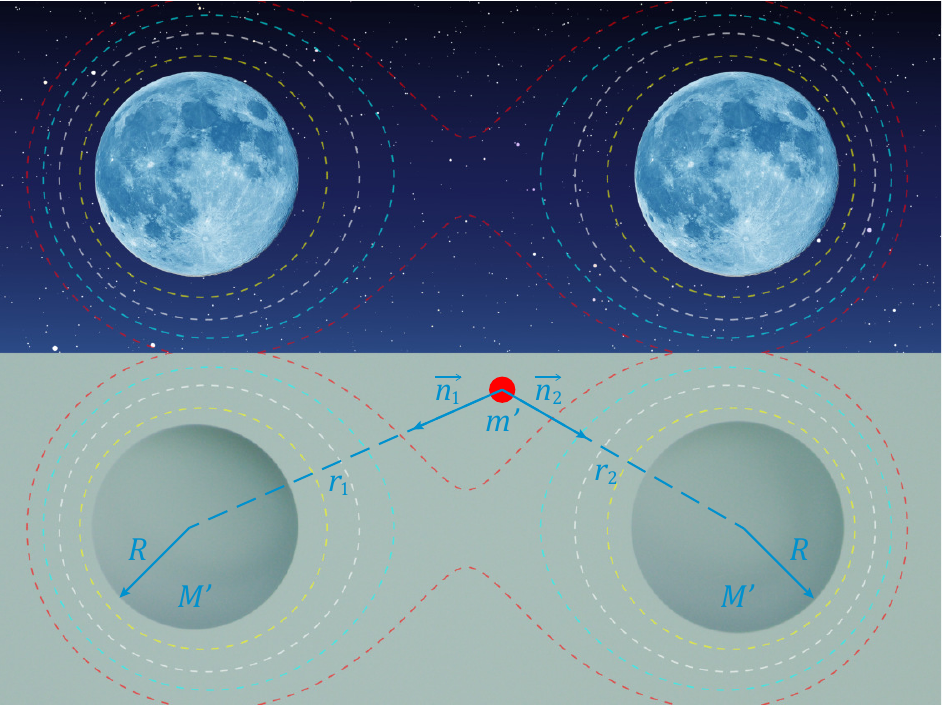}
\caption{(Top) Gravitational equipotential lines between 
two celestial bodies. 
From outside to inside, 
the gravitational potentials represented are $E_0$, 
$1.09E_0$, $1.20E_0$ and $1.38E_0$, respectively. 
(Bottom) Humidity equipotential lines between 
two 10 wt\% PG droplets on a high energy substrate.
From outside to inside, 
the humidity potentials represented are $E^{\prime}$, 
$1.09E^{\prime}$, $1.20E^{\prime}$ 
and $1.38E^{\prime}$, respectively.}\label{use}
\end{figure*}

Back to the previous problem, 
now we can verify Eq. \ref{force_formular_2} with kinematic data 
shown in Fig. \ref{display_all}. 
Balancing drag force $F_d = 3\pi\mu R U l_n/\theta$ with driving force 
$F$ in Eq. \ref{force_formular_2}, 
we get {\color{black}$U = - \gamma_d m_p \lambda R^2 \theta \phi_\infty 
(1 - \phi_\infty)/[3 \mu l_n (L + 2R)^2]$}. 
Combining with Eq. \ref{diffstart_formular}, 
the prefactor in Eq. \ref{cube_formular} finally reads 
\begin{equation}
{\color{black}k = 2 \gamma_d m_p \lambda R^2 \theta \phi_\infty (1 - \phi_\infty) / \mu l_n}.
\label{kt}
\end{equation}

Red dots in Fig. \ref{drag}b show a good agreement between the 
prefactors $k_t$ calculated from Eq. \ref{kt}, 
and $k_e$ measured experimentally from Fig. \ref{display_all}d, 
for $\phi_\infty \sim$ 60\%. 
If we take a linear model of $\cos \theta  = m_{l} \phi+b_{l}$ \cite{cira2015vapour}, 
or $\theta = m_t \left(\phi_{e q}-\phi\right)^{1 / 3}$ \cite{karpitschka2017marangoni}, 
the calculated prefactors $k_t$ marked by gray and blue dots respectively in 
Fig. \ref{drag}b remain consistent with $k_e$.
{\color{black}For harsher experimental conditions with $\phi_\infty <$ 25\% or $>$ 75\%, 
$k_e / k_t$ calculated from all three models deviate from 1, 
but still spread mainly between 0.5 and 1.5 (Fig. \ref{drag}c). 
{\color{black}The deviation possibly originates from the use 
of $\theta = f(\phi_\infty)$ to predict the influence of local 
relative humidity on $\theta$ (Fig. \ref{humidity_all}c). 
A more precise model of $\theta = f(\phi)$ may worth pursuing, 
nevertheless}, 
Fig. \ref{display_all}d has demonstrated that Eq. \ref{cube_formular} 
holds for all tested experimental conditions with $\phi_\infty \in [19.1\%, 81.1\%]$, 
indicating the inverse-square nature of $F$ is beyond reasonable doubt}.

Next we investigate the universal applicability of the inverse-square law force $F$. 
The first test is for two droplets of unequal size. 
Our experimental results (Fig. \ref{extension_mod}a, 
also see Figs. S{\color{black}9-10} in Supplementary Information) 
show that the kinematic relationship satisfies
\begin{equation}
    (L + R_1 + R_2)^3 - (R_1 + R_2)^3 = k' (\tau - t),
    \label{unequal}
\end{equation}
where $R_1$, $R_2$ denotes the spreading radius of the two 
droplets, respectively.
Equation \ref{unequal} has essentially the same form as 
Eq. \ref{cube_formular}, implying the inverse-square 
nature of the force. 
Previously, we have shown the feasibility of using the 
value of $\it \Gamma$ at center of ``target'' droplet 
to calculate $F$. 
Similarly, resultant force on ``target'' droplet here reads 
{\color{black}$F_{12} = \pi R_2^2\mathit{\Gamma} = 
- \lambda m_p \gamma_d \phi_\infty \pi R_1 R_2^{2} 
(1 - \phi_\infty) / (L + R_1 + R_2)^2$}, 
if we take the previously used empirical model $\cos\theta = m_p\phi^2/2 + b_p$. 
Meanwhile, the ``target'' droplet will exert a force on 
the ``source'' droplet reads 
{\color{black}$F_{21} = \pi R_1^2\mathit{\Gamma} = 
- \lambda m_p \gamma_d \phi_\infty \pi R_2 R_1^{2} 
(1 - \phi_\infty) / (L + R_1 + R_2)^2$}. 
It is evident that for different sized droplets, 
$F_{12} \not= F_{21}$, 
indicating they are not action and reaction forces. 
A complete derivation of the force here with accuracy analysis 
can be found in Supplementary Information section {\color{black}VII}.  
For 0.37 $< R1 / R2 < 2.70$, 
the error to above $F_{12}$ and $F_{21}$ will be less than 20\%. 
Meanwhile, the calculated $k_e / k_t$ is nearly equal to 1 
for moderate droplet size ratio but increases to 1.7 
for very unequal droplet sizes (Fig. \ref{extension_mod}a). 
Anyhow, Eq. \ref{unequal} indicates with great certainty the 
inverse-square nature of the force, 
which is profoundly important than the concrete form of it. 

Figure \ref{extension_mod}b implies that the inverse-square 
form of $F$ holds when concentration of PG water solution changes. 
After being redrawn, 
data extracted from previous studies of two pure {\it n}-Hexane 
droplets on a solid substrate 
(Fig. \ref{extension_mod}c) \cite{sadafi2019vapor}, 
and of floating binary droplets on a liquid free surface 
(Fig. \ref{extension_mod}d) \cite{liu2018vapor}, 
indicate that the inverse-square force also applies to such 
pure-liquid-droplet system and thermocapillary system.
For the latter, nonuniform vapor density produces a temperature 
gradient over the cap of the floating droplet and subsequently 
a thermocapillary stress contributing to droplet motion \cite{liu2018vapor}. 
{\color{black}The linear dependence of thermal stress on temperature, 
combining with the inversely proportional decay of temperature in the space, 
guarantees the inverse-square form of the force, 
evidenced by $(L_0+2R)^3-(L+2R)^3\propto t$, 
where $L_0$ is the initial edge-to-edge distance between droplets 
(see detailed derivation in Supplementary Information section VIII)}.
To obtain a straight line through the origin, 
time $\it t$ rather than $\tau - t$ in Fig. \ref{extension_mod}d 
is used to present the kinematic relationship. 
The reason lies in that the thermally affected zones of two
droplets will overlap as distance $L$ keep decreasing, 
which in turn interferes droplet motion, 
evidenced by data deviation from prediction for $t >$ 50 s. 

Next, the key factors in the evaporation-induced 
inverse-square force will be discussed from a perspective 
analogous to the law of universal gravitation.
For gravitational force, 
when a target object with mass $m$ perceives the 
inverse proportion decay gravitational potential 
of a source object $E=-GM/r$, 
it will be attracted by the inverse-square force $F_m=GMm/r^2$, 
where $G$ is the gravitational constant and $M$ is the mass of source object. 
{\color{black}Analogically, we define a humidity potential 
$E^{\prime} = -G^{\prime} M^{\prime} / r = -\gamma_d m_p\phi_\infty 
\lambda(1 - \phi_\infty)R_1 / r$ which also decays inversely proportional to distance, 
where $G^{\prime} = d\gamma / d\phi = \gamma_d m_p \phi_\infty$
is a constant denoting the derivative of surface tension 
to environment humidity, 
$M^{\prime} = \lambda(1 - \phi_\infty) R_1$ represents 
the decay coefficient of humidity away from the source droplet, 
$m^{\prime} = \pi R_2^2$ represents the wetting area of 
the target droplet and $r$ denotes the distance between the 
centers of the two droplets, respectively. 
In an identical coordinate system as the universal gravitation (Fig. \ref{use}), 
when a target droplet senses $E^{\prime}$, 
it will be attracted by a force 
\begin{equation}
   F = \frac{\gamma_d m_p \phi_\infty \lambda \pi R_1 R_2^2 (1 - \phi_\infty)} 
{(L+R_1+R_2)^2} = G^{\prime} \frac{M^{\prime} m^{\prime}}{r^2}. 
   \label{comparisonF}
\end{equation}
It is worth noting that compared with previous force expression, 
$F$ in Eq. \ref{comparisonF} lacks a negative sign, 
as positive directions of force in the two coordinates in 
Fig. \ref{humidity_all}a and Fig. \ref{use} are opposite.} 

In nature, an important manifestation of the law of 
universal gravitation and Coulomb’s law is the satisfaction 
of the superposition principle. 
The superposition of the humidity field is not that straightforward 
as there is an upper limit for humidity, which is 1. 
A detailed description of calculation and fitting of the 
total humidity distribution from two ``source'' droplets 
can be found in Supplementary Information section {\color{black}IX}. 
For simplicity, considering two droplets of the same spreading 
radius $R$, their total humidity is expressed as 
$\phi = \lambda'(1-\phi_\infty)R /r_1 + 
\lambda'(1-\phi_\infty)R / r_2 + \phi_\infty$, 
where $\lambda'$ is a fitting parameter dependent on the distance 
between two droplets (Fig. S12), 
$r_1$ and $r_2$ are distances between ``target'' and the two ``source'' 
droplets (Fig. \ref{use}). 
Not surprisingly, the humidity from multiple droplets satisfies the 
superposition principle.  
For an ideal small ``target'' droplet which does not affect $\phi$, 
the resultant force exerting on it reads
\begin{equation}
   \vec{F}=G^{\prime}\frac{M^{\prime}m^{\prime}}{r_1^2}\vec{n}_1+G^{\prime}\frac{M^{\prime}m^{\prime}}{r_2^2}\vec{n}_2, 
   \label{sumF}
\end{equation}
where $G^{\prime}=\gamma_d m_p \phi_\infty$, $M^{\prime}=\lambda'(1-\phi_\infty)R$, 
$m^{\prime}=\pi a^2$, 
$a$ is the radius of the ideal small ``target" droplet, 
$\vec{n}_1$ and $\vec{n}_2$ are the unit vectors pointing from 
center of ``target'' to the center of two ``source'' droplets, respectively. 
The resultant force has an equivalent form as the superimposed gravitational force.

To conclude, we discover a new inverse-square law force between vapor-mediated droplets. 
We believe it will apply to broader systems beyond what have been investigated here, 
and will be highly beneficial to droplet research and applications. 
With a clear-cut ``propagator'', 
the problem studied here could provide new perspectives 
for a better understanding of inverse square laws. 

This study was supported by the National Natural Science Foundation 
of China (11972339, 11772327, 11932019, 11621202), 
the Fundamental Research Funds for the Central Universities (WK2090000023), 
and the Strategic Priority Research Program of the Chinese Academy of Sciences (XDB22040403).

	\clearpage
	\setcounter{figure}{0}
\setcounter{table}{0}
\setcounter{equation}{0}
\renewcommand{\figurename}{FIG. S\!\!}
\renewcommand{\tablename}{TAB. S\!\!}
\renewcommand\theequation{S\arabic{equation}}

\begin{@twocolumnfalse}
\begin{center}
	{\bf Supplementary Information to: \\ 
	The Inverse-Square Law Force between Vapor-Mediated Droplets}
\end{center}
\end{@twocolumnfalse}

\section{Materials and experiments.}

\begin{table}[htbp]
	\begin{tabular}{lccc}
		\hline
		\hline
		& Density    & Surface  & Dynamic      \\
		& $\rho$     & Tension $\gamma$  & Viscosity $\mu$ \\                    
		& (kg m$^{-3}$) & (mN m$^{-1}$)   & (mPa$\cdot$s) \\ \hline
		air                      &1.20        &---       &0.018      \\
		water                    & 998        & 72.7     & 1.0       \\
		PG                       & 1038          & 35.7        & 56.0        \\
		10 wt\% PG aqueous solution & 1005      & 59.2        & 1.4       \\
		20 wt\% PG aqueous solution & 1014          & 55.7        & 2.0       \\
		40 wt\% PG aqueous solution & 1032          & 47.8        & 4.6       \\
		60 wt\% PG aqueous solution & 1043          & 38.7        & 9.4       \\ \hline
		\hline
	\end{tabular}
	\caption{\label{tab:1}
		Properties of the liquids used in the experiments. All values were measured at room temperature of 20 $^{\circ}$C.}
\end{table} 

\begin{figure}[htbp]
	\includegraphics[width=0.75\linewidth]{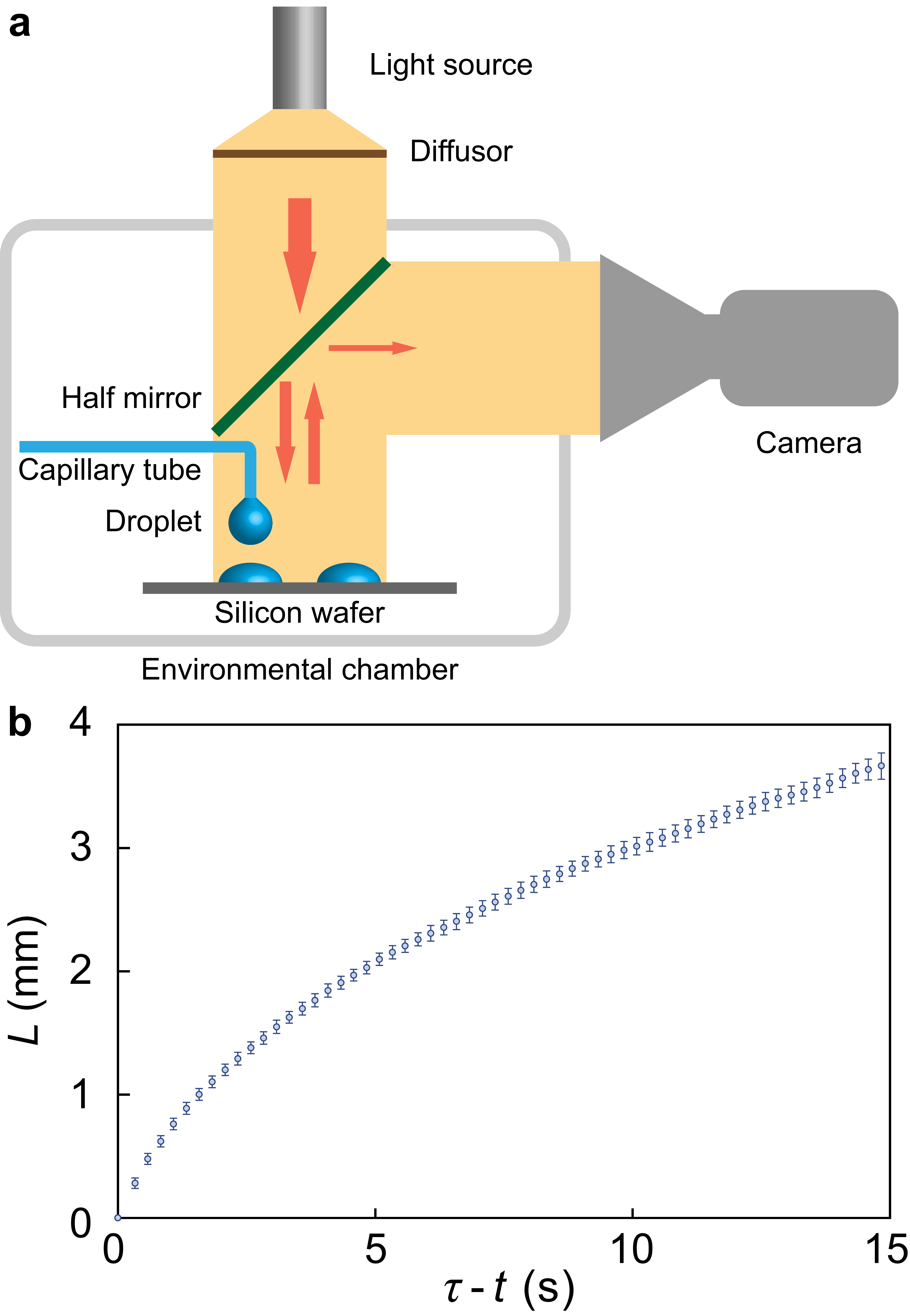}
	\caption{\label{fig:s1}
		(a) Sketch of the experimental configuration, showing the capillary micro-tube used to pinch off the droplet. (b) Mean distance between droplets as a function of time before contact for two moving 10 wt\% PG droplets. 
		$R$ = $1.37 \pm 0.02$ mm and $\phi_\infty$ = $51.3\% \pm 2.0$\%. 
		The error bars represent the standard deviation of 10 experiments.}
\end{figure}

\section{Humidity distribution around a sessile droplet.}
The evaporation process of droplets is a diffusion process, and the relative air humidity $\phi$ satisfies the diffusion equation,
\begin{equation}
	\frac{\partial \phi}{\partial t}=D \nabla^{2} \phi,
	\label{eqn:1}
\end{equation} 
\noindent where $D=26.1\ \rm{mm^2\ s^{-1}}$ is the diffusion coefficient \cite{Hu2002}. In our experiments, the characteristic length $R \sim 1\ \rm{mm}$, characteristic time is $t_c \sim 1\ \rm{s}$, leading to a diffusion time scale $t_d=R^2/D=0.04\ \mathrm{s}\ll \mathit{t_c}$. Therefore, droplet motion can be treated as a quasi-static process, 
during which Eq. \ref{eqn:1} can be simplified as Laplace's equation $D \nabla^{2} \phi=0$. 
In spherical coordinates it reads 
\begin{equation}
	\frac{1}{r^{2}} \frac{\partial}{\partial r}\left(r^{2} \frac{\partial \phi}{\partial r}\right)=0,
	\label{eqn:2}
\end{equation}
\noindent 
$\phi(R)=1,\ \phi(\infty)=\phi_{\infty}$, 
where $R$ is droplet radius, $\phi_{\infty}$ is the ambient humidity in the far field. 
The centrosymmetric solution for $\phi$ around a spherical droplet with an infinity boundary reads 
\begin{equation}
	\phi(r)=\frac{\left(1-\phi_{\infty}\right) R}{r}+\phi_{\infty}.
	\label{eqn:3}
\end{equation}\par
For evaporation from a sessile droplet on a solid substrate, 
we can write the droplet profile for a spherical cap
\begin{equation}
	h(r)=\sqrt{R^2/\sin^2\theta-r^2}-R/\tan\theta, 
\end{equation}
and the boundary conditions 
\begin{align}
	&r<R,\ z=h(r):\ \phi=1; \notag \\
	&r>R,\ z=0:\ J=0; \notag\\
	&r=\infty,\ z=\infty:\ \phi=\phi_{\infty}, 
\end{align}
\noindent 
where $J$ is the evaporation flux, 
$\theta$ is the contact angle, 
$h$, $r$ and $z$ are geometric parameters, 
as shown in Fig. S2a. 
In toroidal coordinate, 
the Laplace equation $\nabla^2\phi=0$ has an analytical solution 
\begin{align}
	&\frac{\phi-\phi_\infty}{1-\phi_\infty}=\sqrt{2\cosh\alpha-2\cos\beta}\cdot\notag\\ 
	&\int_0^\infty\frac{\cosh{(\theta\omega)}}{\cosh{(\pi\omega)}}
	\frac{\cosh{[(2\pi-\beta)\omega]}}{\cosh{[(\pi-\theta)\omega]}}P_{-1/2+i\pi}(\cosh\alpha)d\omega, 
	\label{analytical_solution}
\end{align}
\noindent 
where $P_{-1/2+i\pi}(\cosh\alpha)$ is the Legendre function of the first kind. 
\begin{align}
	P_{-1/2}& _{+i\pi}(\cosh\alpha)\notag\\
	&=\frac{2}{\pi}\coth{(\pi\omega)}\int_\alpha^\infty\frac{\sin(\omega t)}{\sqrt{2\cosh t-2\cosh\alpha}}dt.
\end{align}
\noindent 
The toroidal coordinates $\alpha$ and $\beta$ (Fig. S2b) are related to the cylindrical coordinates $r$ and $z$ by
\begin{align}
	r =& \frac{R\sinh\alpha}{\cosh\alpha-\cos\beta}, \notag\\
	z= & \frac{R\sin\beta}{\cosh\alpha-\cos\beta}.
	\label{conversion}
\end{align}

\begin{figure}[tbp]
	\includegraphics[width=\linewidth]{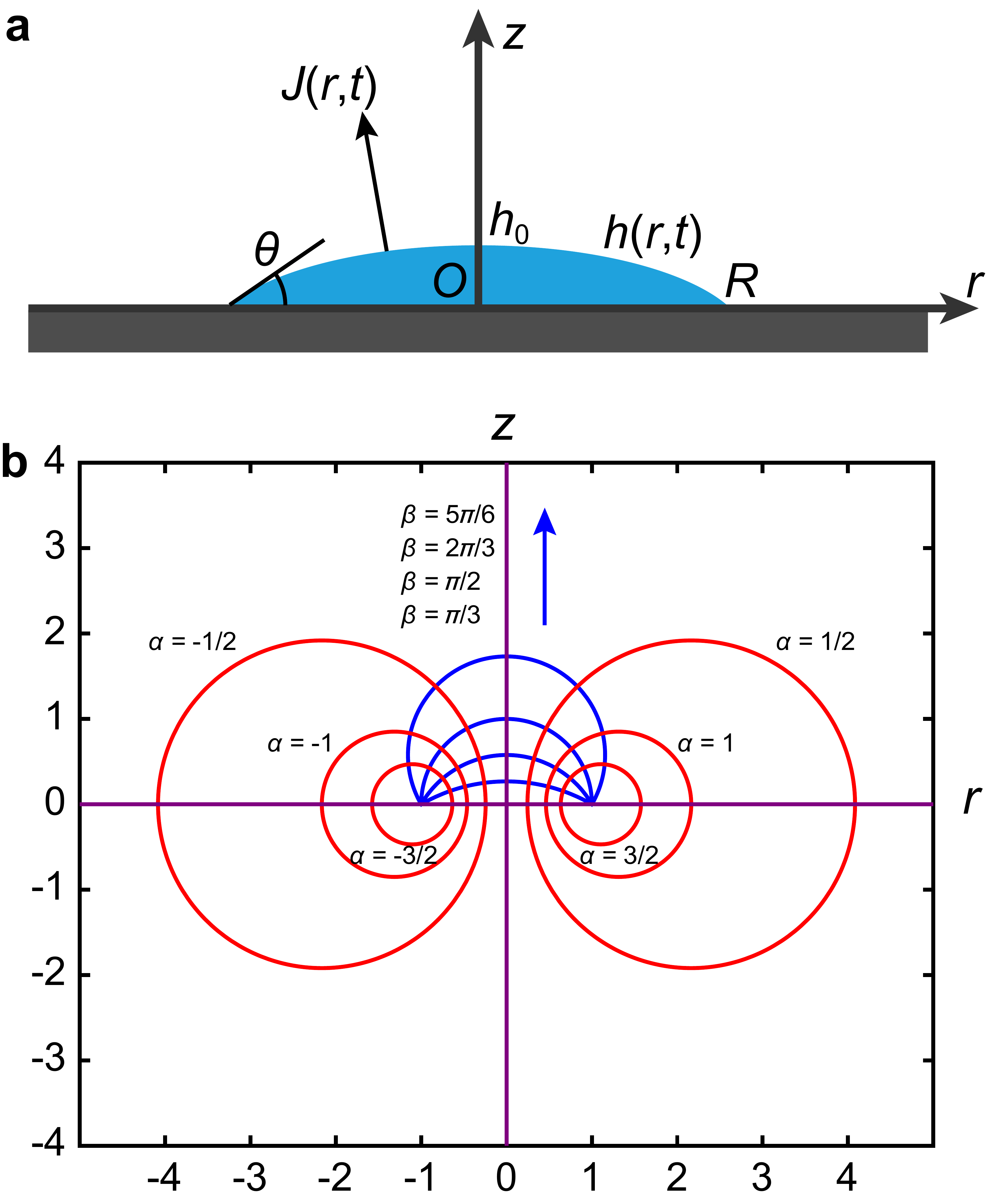}
	\caption{\label{fig:s2}Schematic showing the evaporation of 
		a sessile droplet in (a) cylindrical coordinate system and 
		(b) toroidal coordinate system.}
\end{figure}

\noindent For a thin liquid film with $\theta$ = 0, 
Eq. \ref{analytical_solution} is reduced to 
\begin{align}
	&\frac{\phi-\phi_\infty}{1-\phi_\infty}\notag\\
	&=\frac{1}{2}-\frac{1}{\pi}\arctan{\Big[\frac{\cos\beta-\sinh^2{(\alpha/2)}}
		{\cos{(\beta/2)}\sqrt{2\cosh\alpha-2\cos\beta}}\Big]}.
	\label{theta=0}
\end{align}

\noindent For humidity distribution along the solid substrate, 
i.e., $z = 0$, Eq. \ref{theta=0} is further reduced to 
\begin{equation}
	\frac{\phi-\phi_\infty}{1-\phi_\infty}=\frac{1}{2}-\frac{1}{\pi}\arctan{\Big[\frac{1-\sinh^2{(\alpha/2)}}
		{\sqrt{2\cosh\alpha-2}}\Big]}.
	\label{D13}
\end{equation}

\begin{figure}[tbp]
	\includegraphics[width=\linewidth]{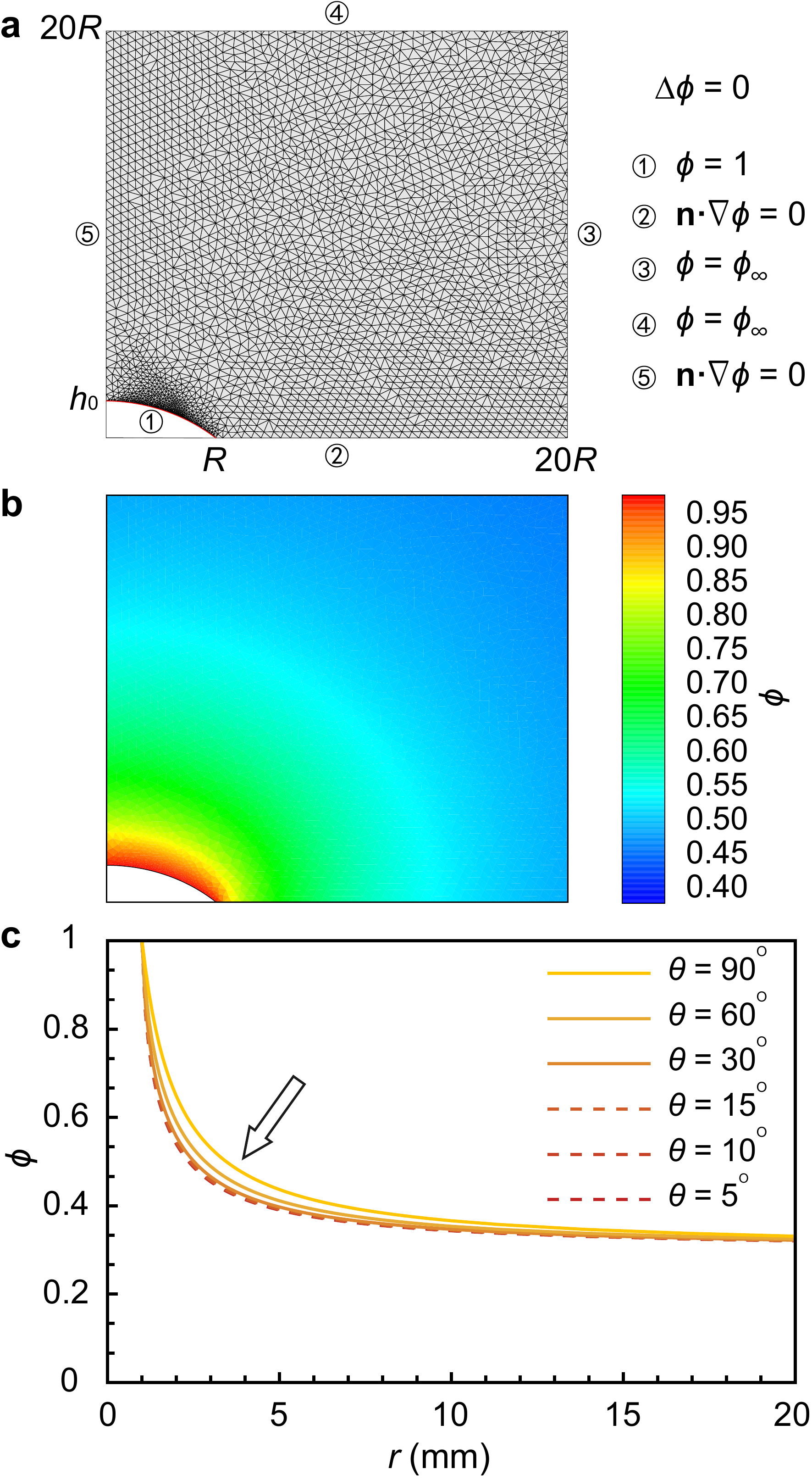}
	\caption{\label{fig:s3}
		(a) Mesh partition and boundary conditions for finite element calculation of $\phi$. 
		(b) Calculated humidity field of an evaporating sessile droplet. 
		$\phi_{\infty} = 30\%$, $\theta = 38^{\circ}$.
		(c) Dependence of $\phi$ on $\theta$. 
		Different curves of $\phi$ converge to a single curve 
		when $\theta$ $\leq$ 30$^{\circ}$. 
		Arrows marks the direction of convergence.}
\end{figure}
\noindent From Eq. \ref{conversion} we get $\sinh{(\alpha/2)}=\sqrt{{1}/{(r^2/R^2-1)}}$. Denoting $\tilde{r}$ as $\tilde{r}=r/R$, the arctangent term in Eq. \ref{D13} can be expressed as 
\begin{align}
	f(\tilde{r})=&\arctan{\Big[\frac{1-\sinh^2{(\alpha/2)}}{\sqrt{2\cosh\alpha-2}}\Big]}\notag\\
	=&\arctan{\Big[\frac{1-\sinh^2{(\alpha/2)}}{2\sinh{(\alpha/2)}}\Big]}\notag\\
	=&\arctan{\Big[\frac{1-{1/(\tilde{r}^2-1)}}{2\sqrt{{1}/{(\tilde{r}^2-1)}}}\Big]}.
	\label{fr}
\end{align}

\noindent The asymptotic expansion of $f(\tilde{r})$ at infinity reads
\begin{equation}
	f(\tilde{r})=\frac{\pi}{2}-\frac{2}{\tilde{r}}-\frac{1}{3\tilde{r}^3}-\frac{3}{20\tilde{r}^5}+O\Big[\frac{1}{\tilde{r}}\Big]^6. 
\end{equation}
\noindent Consider the correction to the first-order term by the higher-order term with least square fit, 
we get 
\begin{equation}
	f(\widetilde{r})=\frac{\pi}{2}-\frac{2.2}{\tilde{r}}.
\end{equation}
\noindent Eventually, the humidity distribution of sessile droplet along the substrate reads
\begin{equation}
	\phi(r)=\frac{\lambda\left(1-\phi_{\infty}\right) R}{r}+\phi_{\infty}, 
	\label{eqn:4}
\end{equation}
\noindent where $\lambda=2.2/\pi\approx0.7$.

To verify our inference, 
Finite element method software FreeFem++ was also used to 
solve the Laplace equation. 
The computational domain and numerical results are shown 
in Fig. S\ref{fig:s3}. 
Fig. S3c shows the dependence of $\phi$ on $\theta$. 
It is shown that different curves of $\phi$ collapse into 
a single curve when $\theta <$ 30$^{\circ}$. 
For our experiments where $\theta\in[5^{\circ},15^{\circ}]$, 
the influence of contact angle on humidity distribution 
is negligible. 
Therefore, we could use Eq. S10 or Eq. S14 to estimate 
humidity distribution $\phi$. 
In the main text, a comparison among theoretical result, asymptotic solution, and numerical result is shown in Fig. 2b. 

\section{Dependence of contact angle on humidity.}
Different models have been used to predict humidity's effects 
on the contact angle of a sessile droplet on a solid substrate. 
For example \cite{Cira2015,karpitschka2017}, 
\begin{align}
	\label{eqn:5}
	\cos \theta &=m_{l} \phi+b_{l}, \\
	\theta &=m_{t}\left(\phi_{e q}-\phi\right)^{1 / 3},
\end{align}
and the model we propose in the main text, 
\begin{equation}
	\cos \theta =\frac{1}{2}m_{p} \phi^2+b_{p},
	\label{eqn:6}
\end{equation}\par
\noindent where $\phi_{eq}$ is the equilibrium relative humidity 
when the droplet spreads completely, 
and $m_l$, $b_l$, $m_t$, 
$m_p$, $b_p$ are fitting parameters. 

\begin{figure}[t]
	\includegraphics[width=\linewidth]{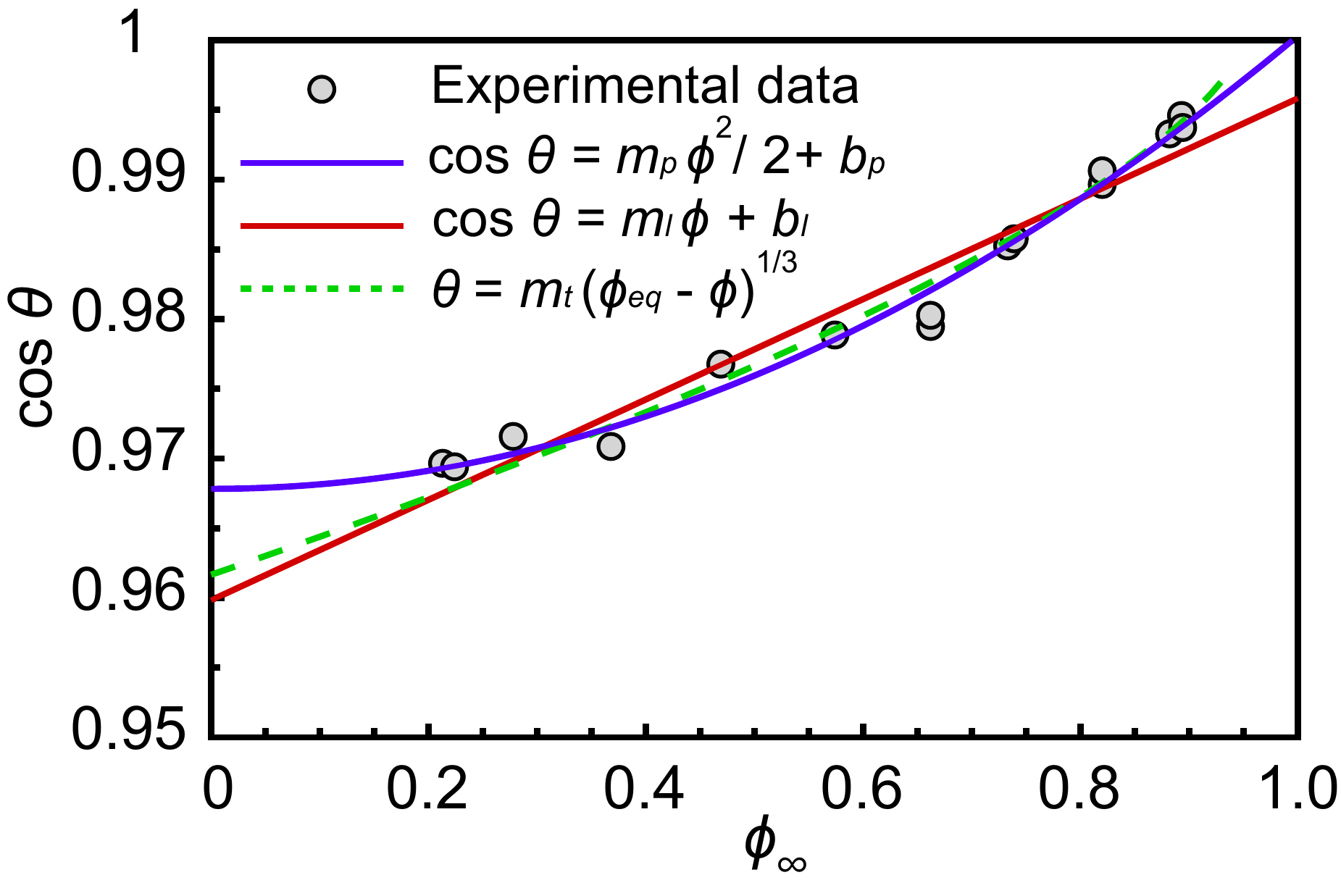}
	\caption{\label{fig:s5}
		Contact angle as a function of $\phi_{\infty}$ 
		for a 10 wt\% PG droplet on a 
		hydrophilic silicon surface.
		The fitting coefficients are $m_p=0.0.65$ and $b_p=0.97$, 
		$m_l=0.036$ and $b_l=0.96$, $m_t=0.28$ and $\phi_{eq}=0.95$, 
		for three different models, respectively.}
\end{figure}
A comparison between experimental data and results obtained 
from three different models is shown in Fig. S\ref{fig:s5}, 
where $m_p=0.0.65$ and $b_p=0.97$, 
$m_l=0.036$ and $b_l=0.96$, 
$m_t=0.28$ and $\phi_{eq}=0.95$, respectively. 

\section{Driven force between evaporating sessile droplets.}
Microscopic observation of the binary droplets revealed 
a thin film extending tens of microns from the edge of 
the bulk droplet \cite{Cira2015}. 
The Young equation indicates that $\gamma=\gamma_d\cos\theta$, 
where $\gamma$ is surface tension along the periphery of 
the ``target'' droplet, and $\gamma_d$ is surface tension of droplet. 
As mentioned above, contact angle is a function of humidity 
$\cos\theta=f(\phi)$. 
To simplify the analysis, $\cos \theta$ is expanded linearly, 
\begin{equation}
	\cos \theta=f^{\prime}\left(\phi_{\infty}\right)\left(\phi-\phi_{\infty}\right)+f\left(\phi_{\infty}\right)=m \phi+b,
	\label{eqn:7}
\end{equation}
\noindent where $m=f^{\prime}(\phi_{\infty})$ and 
$b=-f^{\prime}(\phi_{\infty})\phi_{\infty}+f(\phi_{\infty})$. 
For the case shown in Fig. 2b in the main text, 
the calculated relative error at distance $r = 2R$ 
(corresponding to $L = 0$) is 1.49\%, 
verifying the feasibility of the linear approximation. 
Resultant force exerted on the ``target" droplet is  
\begin{align}
	F &=\int_{0}^{2 \pi} \gamma \cos \varphi R d \varphi\notag \\
	&=\int_{0}^{2 \pi} \gamma_{d} \cos \theta \cos \varphi R d \varphi\notag \\
	&=\int_{0}^{2 \pi} \gamma_{d}(m \phi+b) \cos \varphi R d \varphi\notag \\
	&=\gamma_{d}m_p \phi_{\infty}  R \int_{0}^{2 \pi} \phi \cos \varphi d \varphi\notag\\
	&=\gamma_{d}m_p \phi_{\infty}  R \int_{0}^{2 \pi}\left[\frac{\lambda\left(1-\phi_{\infty}\right) R}{r}+\phi_{\infty}\right] \cos \varphi d \varphi\notag \\
	&=\gamma_{d} m_p   \phi_{\infty}\lambda R^{2}\left(1-\phi_{\infty}\right) \int_{0}^{2 \pi} \frac{\cos \varphi}{r} d \varphi. 
	\label{eqn:8}
\end{align}
\noindent Introducing a relatively small variable $\epsilon_1 = R /(L+2 R)$ where $0<\epsilon_1 < 1/2$, 
\begin{align}
	\frac{1}{r} =&\frac{1}{\sqrt{(L+2R+R\cos\varphi)^2+(R\sin\varphi)^2}}\notag \\
	= &\frac{1}{L+2R}\frac{1}{\sqrt{1+2\epsilon_1\cos{\varphi}+\epsilon_1^2}}\notag \\
	=&\frac{1}{L+2R}\sum_{n=0}^{\infty}{P_n(-\cos\varphi)\epsilon_1^n}\notag\\
	\approx & \frac{1}{L+2R}\Big[1-\epsilon_1\cos\varphi+\frac{\epsilon_1^2}{2}(3\cos^2\varphi-1)\notag\\
	&+  \frac{\epsilon_1^3}{2}(-5\cos^3\varphi+3\cos\varphi)+o(\epsilon_1^3)\Big], 
\end{align}
\noindent where $P_n$ is the Legendre polynomial of order $n$. 
Thereby we have 
\begin{align}
	\int_{0}^{2 \pi}& \frac{\cos \varphi}{r} d \varphi\notag\\
	\approx&\frac{1}{L+2 R} \int_{0}^{2 \pi} \cos \varphi[1-\epsilon_1 \cos \varphi +\frac{\epsilon_1^{2}}{2}\left(3 \cos ^{2} \varphi-1\right)\notag \\
	&+\frac{\epsilon_1^{3}}{2}\left(-5 \cos ^{3} \varphi+3 \cos \varphi\right)] d \varphi\notag \\
	=&-\frac{\pi R}{(L+2 R)^{2}}-\frac{3 \pi R^{3}}{8(L+2 R)^{4}}. 
	\label{eqn:10}
\end{align}
\noindent Ignoring higher-order terms, the resultant force of third-order accuracy can be expressed as
\begin{equation}
	F=-\frac{\gamma_{d} m_p \phi_{\infty} \lambda \pi R^{3}\left(1-\phi_{\infty}\right)}{(L+2 R)^{2}}.
	\label{eqn:11}
\end{equation}
\noindent The residual is $3\epsilon_1^2/8<5\%$ when $L > R$.

\section{Drag of a moving droplet.}
\begin{figure}[htbp]
	\includegraphics[width=0.8\linewidth]{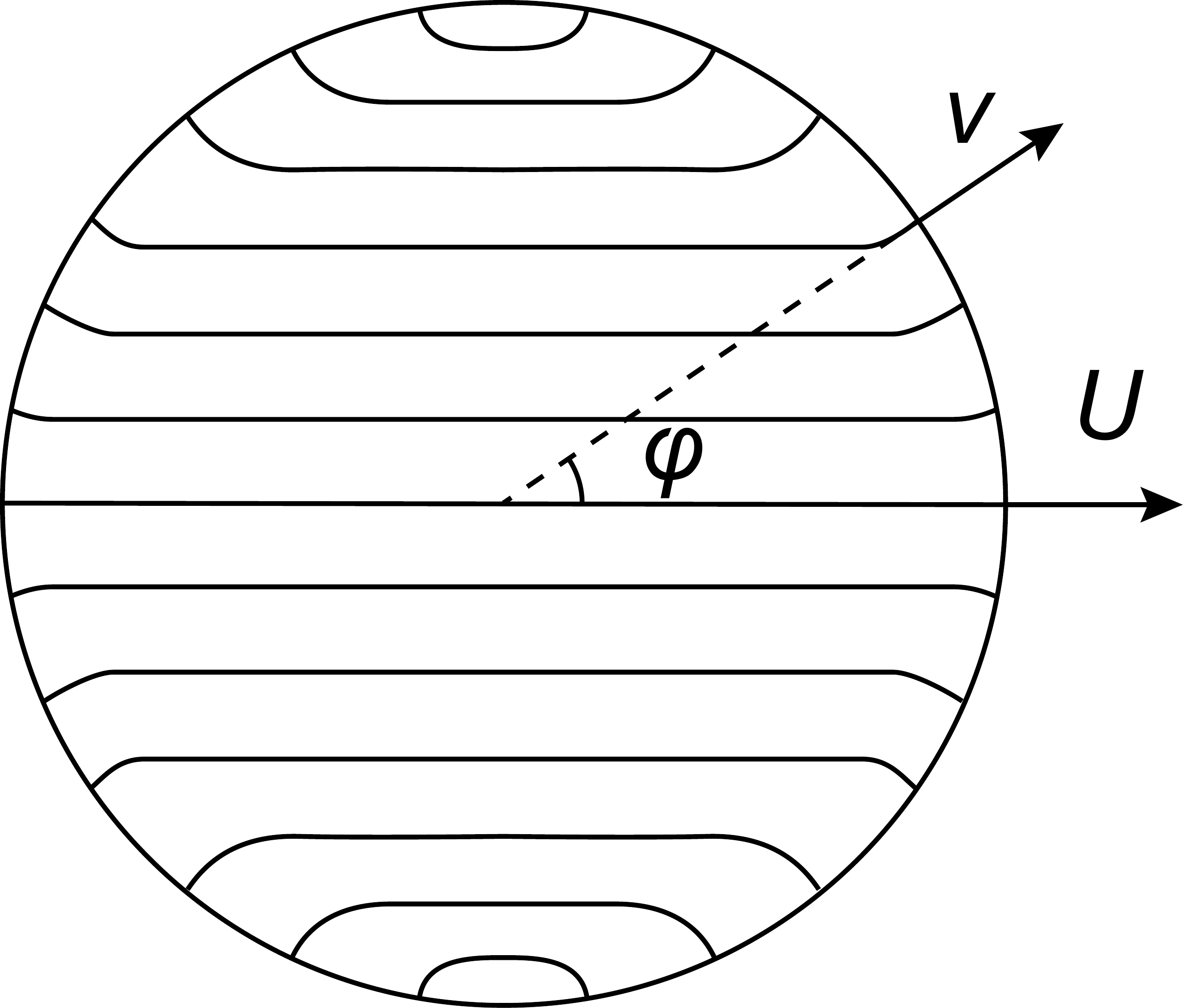}
	\caption{\label{fig:s6}
		The local velocity of the droplet liquid is perpendicular to 
		the contact line. 
		A copy sketch to fig. 2b from \cite{reyssat2014}.}
\end{figure}

When a droplet moves on the silicon wafer, 
drag $F_d = F_{cl} + F_{bulk}$, 
where $F_{cl}$ and $F_{bulk}$ originates from moving contact line 
and the wetting area, respectively \cite{Bico2001}. 
In the vicinity of contact line, 
previous works show that the local velocity of the droplet liquid 
is perpendicular to the contact line \cite{brochard1989, Rio2005}, as shown in Fig. S\ref{fig:s6}. 
The drag $f$ exerted on per unit contact line length is 
\begin{equation}
	f=\frac{3 \mu l_{n} v}{\theta}=\frac{3 \mu l_{n} U \cos \varphi}{\theta},
	\label{eqn:12}
\end{equation}
\noindent where $\mu$ is dynamic viscosity of liquid, 
$v$ is the local velocity of contact line, 
$\theta$ is dynamical contact angle, $l_n = \mathrm{ln} (l_{max}/l_{min})$ is the logarithmic cut-off coefficient \cite{HUH1971}, 
$l_{max}$ is the macroscopic characteristic length 
and $l_{min}$ is the microscopic characteristic length, respectively. 
Generally, $l_n\approx 15$ for droplet moving on a dry surface \cite{hoffman1975} and $l_n\approx 5$ for droplet moving on a wet surface \cite{Bico2001}. 

For constant $\theta$ during droplet motion, 
the drag originating from contact line can be derived by integrating Eq. \ref{eqn:12} along the contact line, 
\begin{align}
	F_{cl}&=\int_{0}^{2 \pi} f \cos \varphi R d \varphi\notag\\
	&=\int_{0}^{2 \pi} \frac{3 \mu l_{n} R U \cos ^{2} \varphi}{\theta} d \varphi\notag\\ &=\frac{3 \pi \mu l_{n} R U}{\theta}.
	\label{eqn:13}
\end{align}
$F_{bulk}$ can be estimated as 
\begin{equation}
	F_{bulk} \sim \frac{\mu U}{h} \pi R^{2}=\frac{\mu U}{R \theta} \pi R^{2}=\frac{\mu \pi R U}{\theta},
	\label{eqn:14}
\end{equation}
\noindent where $h\approx R\theta$ is the height of the sessile droplet. 
Comparing Eq. \ref{eqn:13} and \ref{eqn:14}, $F_{cl}/F_{bulk}\sim 3l_n \gg O(1)$, therefore drag $F_{bulk}$ is negligible and the 
total drag of droplet reads
\begin{equation}
	F_d = \frac{3 \pi \mu l_{n} R U}{\theta}.
	\label{eqn:15}
\end{equation}

\begin{figure}[htbp]
	\includegraphics[width=\linewidth]{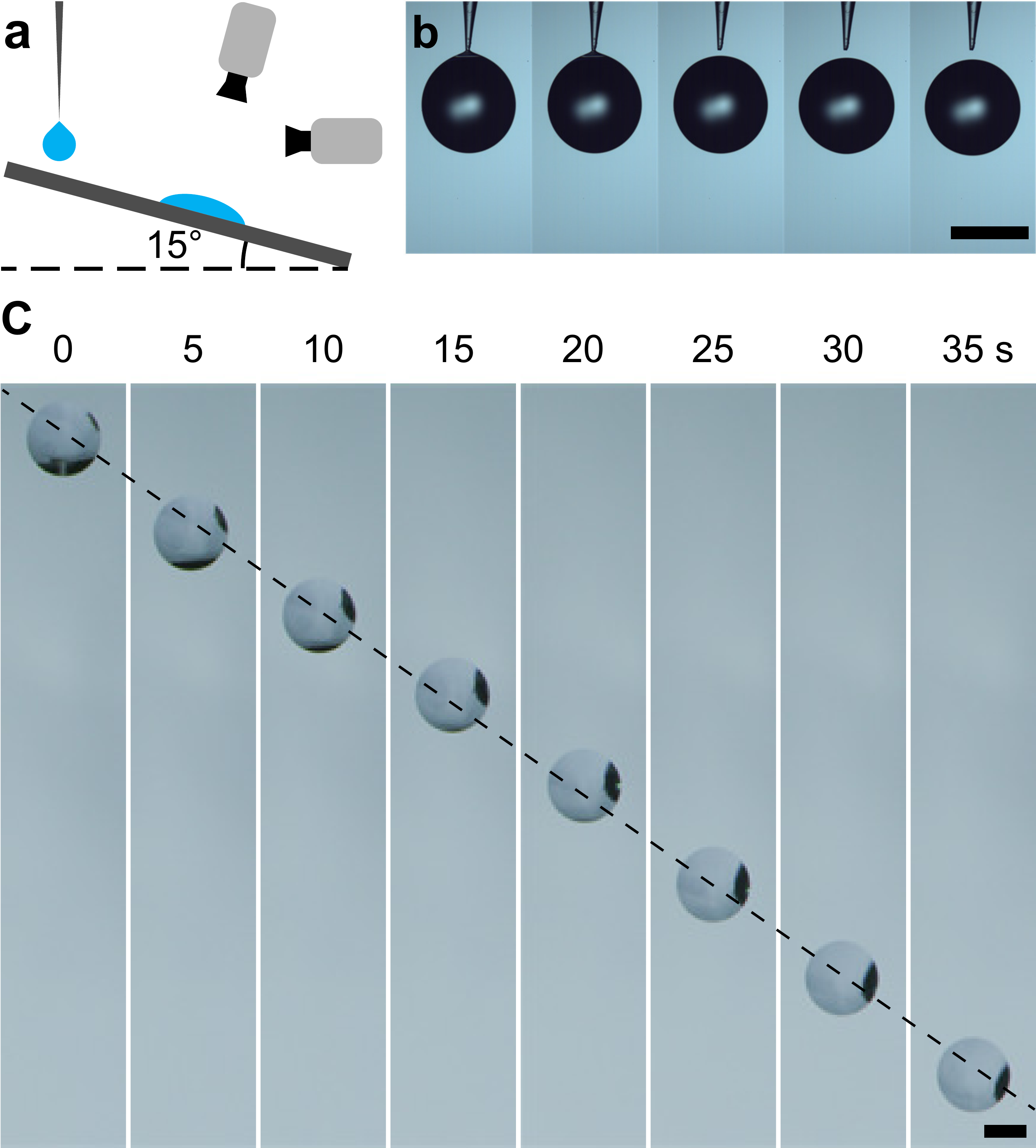}
	\caption{\label{fig:s7}
		(a) The sketch of a droplet sliding along a silicon wafer with an inclination angle of 15$^{\circ}$. 
		(b) The pinch-off of a 10 wt\% PG aqueous droplet from a conical 
		capillary tube. Scale bar is 1 mm. 
		(c) Droplet sliding down along a silicon wafer with an inclination angle of 15$^{\circ}$. 
		The droplet has reached its terminal velocity. 
		scale bar is 2 mm.}
\end{figure}

If the dynamic contact angle changes with droplet velocity, 
the classical Cox-Voinov law gives
\begin{equation}
	\theta^3=\theta_m^3 + 9 l_n Ca,
	\label{eqn:16}
\end{equation}
\noindent where $\theta_m$ is the microscopic contact angle and $Ca=\mu U/\gamma$ is the capillary number. 
Due to the existence of a thin liquid film around the edge 
of droplet in our experiments, we have $\theta_m = 0$, 
and the Cox-Voinov law can be simplified to Tanner's law 
which gives $\theta\sim Ca^{1/3}$. 
Consequently Eq. \ref{eqn:13} and Eq. \ref{eqn:14} will be 
reduced to 
\begin{equation}
	F_{cl}^{\prime}=\frac{3 \pi\mu l_{n}R U}{\theta} \sim \frac{\gamma RC a}{C a^{1 / 3}}=\gamma RC a^{2 / 3}
	\label{eqn:17}
\end{equation}
\noindent and 
\begin{equation}
	F_{bulk}^{\prime} \sim \frac{\pi\mu UR^2}{h} = \frac{\pi\mu U R}{\theta} \sim \gamma R C a^{2 / 3},
	\label{eqn:1000}
\end{equation}
respectively. 
The total drag reads 
\begin{equation}
	F_{d}^{\prime}=F_{cl}^{\prime}+F_{bulk}^{\prime} \sim \gamma R C a^{2 / 3}.
	\label{eqn:18}
\end{equation}

To verify above analysis, 
systematic experiments were conducted for droplet sliding 
down process along a tilt silicon wafer with an inclination 
angle of 15$^{\circ}$. 
The driving force is the gravity force component along the substrate
\begin{equation}
	F_g = \rho \mathit{\Omega} g \sin\mathrm{15}^\circ,
	\label{eqn:19}
\end{equation}
\noindent where $\rho$ is the liquid density, $\it \Omega$ is the droplet volume and $g$ is the gravity acceleration, respectively. 
$\rho$ was measured by a side-view high-speed camera, 
and the droplet sliding process was observed by another high-speed camera perpendicular to substrate (Fig. S\ref{fig:s7}a). 
When the droplet reaches its terminal velocity, 
$F_g$ is balanced by $F_d$. 
Comparing Eq. \ref{eqn:15}, \ref{eqn:18} and Eq. \ref{eqn:19}, 
we obtain two expressions 
\begin{align}
	\rho \mathit{\Omega} g \sin\mathrm{15}^\circ&=\frac{3 \pi \mu l_{n} R U_t}{\theta},\label{eqn:20} \\
	\rho \mathit{\Omega} g \sin\mathrm{15}^\circ&=k_d\gamma R C a_t^{2 / 3},
	\label{eqn:21}
\end{align}
\noindent where $U_t$ is terminal velocity, 
$Ca_t=\mu U_t/\gamma$ and $k_d$ is a fitting coefficient, respectively. 
Feasibility of Eqs. \ref{eqn:20} and \ref{eqn:21} were 
tested in Fig. 3a in the main text. 
The experimental data fits better with Eq. \ref{eqn:15} and 
the best fit gives $l_n = 17.8$.

\section{Effects of droplet deformation on its motion.}
\begin{figure}[htbp]
	\includegraphics[width=\linewidth]{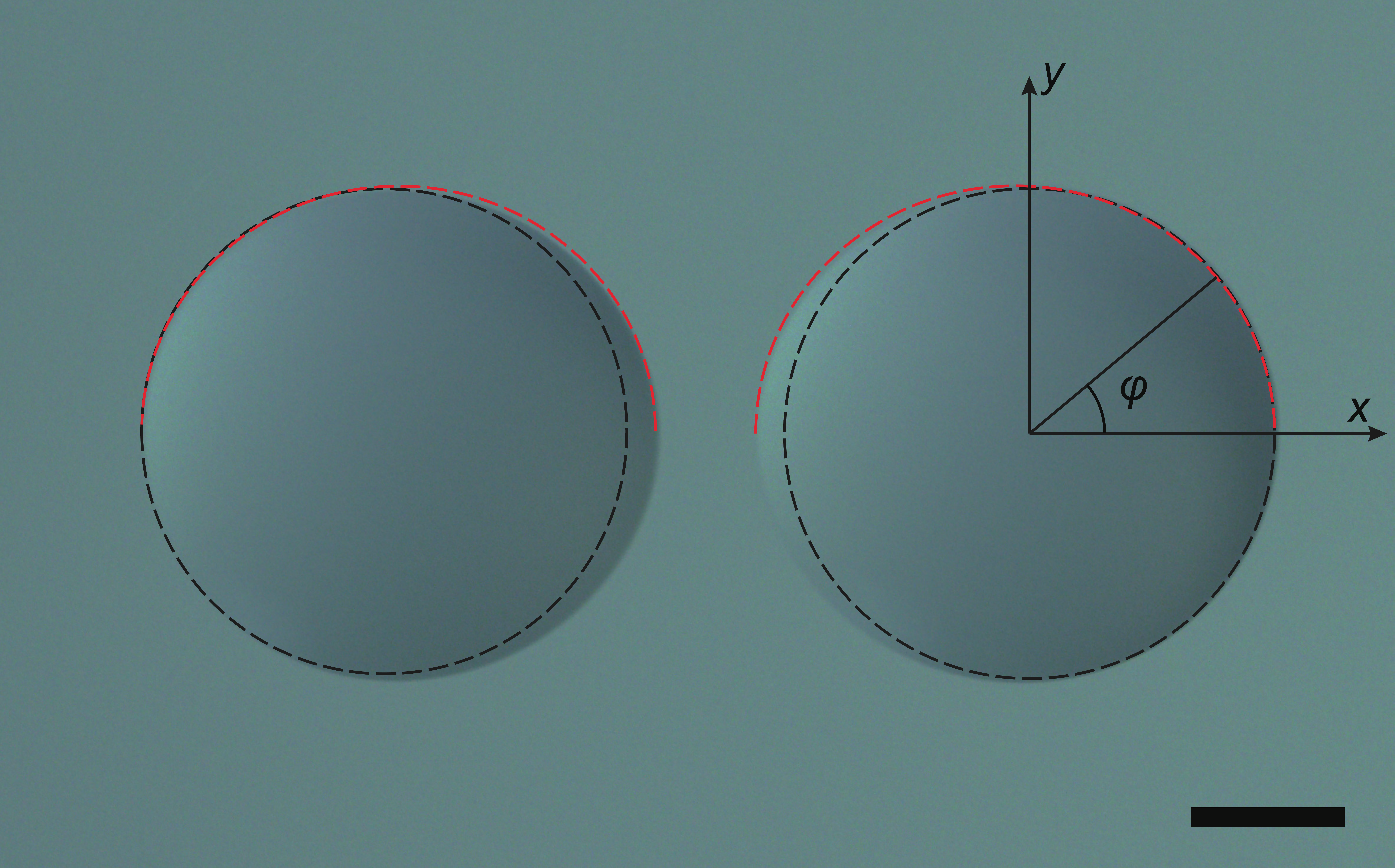}
	\caption{\label{fig:s8}
		Small deformation of adjacent 10 wt\% PG aqueous droplets 0.1 s before contact. 
		The semicircle composed of red broken lines is used to mark the outline of the droplet.
		Scale bar is 1 mm.}
	\label{small}
\end{figure}

Just before contact, 
droplets will undergo a small deformation, 
as shown in Fig. S\ref{small}. 
Below we will investigate its effects on the driving force $F$ that exerted on droplet.  

For a small deformation, 
the drag at the receding contact line keeps constant as
\begin{align}
	F_d^r =& 2\int_0^{\pi/2}f_r\cos\varphi Rd\varphi\notag \\
	= & 2\int_0^{\pi/2}\frac{3\mu l_n RU \cos^2\varphi}{\theta}d\varphi\notag\\
	=& \frac{3\pi\mu l_n RU}{2\theta}, 
\end{align}
where $f_r$ denotes the drag exerted on per unit advancing contact line length. 

At the advancing contact line, 
the deformation of the droplet can be expressed with 
$r=R[1+\epsilon_2 f(\varphi)]$ in polar coordinates (Fig. S\ref{small}) where $\epsilon_2 < 0.1$ is a negligible constant. 
With $x = r(\varphi)\cos\varphi$ and $y = r(\varphi)\sin\varphi$, 
the tangent vector of the advancing contact line can be expressed as
\begin{equation}
	\begin{aligned}
		\textbf{T} =\Big[\frac{dx}{d\varphi},\frac{dy}{d\varphi}\Big]^T, 
	\end{aligned}
\end{equation}
\noindent and the unit tangent vector and the unit normal vector can be expressed as 
\begin{equation}
	\textbf{t}={\textbf{T}}/{||\textbf{T}||},\quad\textbf{n}=\textbf{A}.\textbf{t}
	,\quad\textbf{A}=\begin{bmatrix} 0 & 1 \\
		-1 & 0
	\end{bmatrix},
\end{equation}
where
\begin{align}
	\textbf{t}= \Big[ &-\sin\varphi +\epsilon_2  \cos \varphi  f'(\varphi )+
	\epsilon_2 ^2 [ \sin \varphi  f'(\varphi )^2/2\notag\\
	&{}-f(\varphi )f'(\varphi ) \cos \varphi  ]+o(\epsilon_2 ^2),\notag\\
	& \cos\varphi+\epsilon_2  \sin\varphi f'(\varphi )+\epsilon_2 ^2 [- \cos\varphi f'(\varphi )^2/2\notag\\
	&{}-f(\varphi )f'(\varphi ) \sin \varphi] +o(\epsilon_2 ^2)\Big]^T, \\
	\textbf{n}= \Big[ &\cos\varphi +\epsilon_2 \sin \varphi  f'(\varphi )+ \epsilon_2^2 [-\cos \varphi  f'(\varphi )^2/2\notag\\
	&{}-f(\varphi )f'(\varphi ) \sin \varphi ]+o(\epsilon_2 ^2),\notag\\
	& \sin\varphi-\epsilon_2 \cos\varphi f'(\varphi )+\epsilon_2^2 [- \sin\varphi f'(\psi )^2/2\notag\\
	&{}+ f(\varphi )f'(\varphi ) \cos \varphi  ]+o(\epsilon_2 ^2) \Big]^T.
\end{align}
Denoting the drag per unit advancing contact line length as 
$f_a = 3\mu l_n Un_x/\theta$, and arc length element as 
\begin{align}
	ds&=\sqrt{r^2+r'^2}d\varphi\notag\\
	&=R[1+\epsilon_2 f(\varphi)+\epsilon_2^2f'^2(\varphi)/2+o(\epsilon_2^2)]d\varphi, 
\end{align}
where $n_x$ is the $x$ component of unit normal vector and $r^{\prime}=dr/d\varphi$, 
\noindent the drag at the advancing contact line reads
\begin{align}
	F_d^a=&2\int_{\pi/2}^{\pi} f_an_xds\notag\\
	=  & 2\int_{\pi/2}^{\pi}\frac{3\mu l_n U n_x^2}{\theta}ds\notag\\
	=  & \frac{6\mu l_n U }{\theta}\int_{\pi/2}^{\pi}n_x^2ds\notag\\
	\approx  & \frac{6\mu l_n RU }{\theta}\int_{\pi/2}^{\pi}
	[\cos ^2\varphi + \epsilon_2\sin2\varphi   f'(\varphi )\notag\\
	&+\epsilon_2 \cos^2\varphi f(\varphi)]d\varphi.
\end{align}
\noindent For the integral in the above formula, 
\begin{align}
	I= & \int_{\pi/2}^{\pi}\left[\cos ^2\varphi + \epsilon_2  \sin2\varphi   f'(\theta )+
	\epsilon_2 \cos^2\varphi f(\varphi)\right]d\varphi\notag \\
	= & \frac{\pi}{4}+\epsilon_2\int_{\pi/2}^{\pi} \sin2\varphi   f'(\varphi )d\varphi+
	\epsilon_2 \int_{\pi/2}^{\pi}\cos^2\varphi f(\varphi)d\varphi\notag  \\
	=&\frac{\pi}{4}-2\epsilon_2 \int_{\pi/2}^{\pi}\cos2\varphi  f(\varphi )d\varphi+
	\epsilon_2 \int_{\pi/2}^{\pi}\cos^2\varphi f(\varphi)d\varphi, 
\end{align}
\noindent and the boundary condition $f(\pi/2)=0,\ f(\pi)=1$. Assuming that $f(\pi)$ is a linear function, 
the integral can be reduced to 
\begin{equation}
	I\approx\frac{\pi}{4}-\frac{2\epsilon_2}{\pi}+\epsilon_2\left(\frac{1}{2\pi}+\frac{\pi}{8}\right)=
	\frac{\pi}{4}-\epsilon_2\left(\frac{3}{2\pi}-\frac{\pi}{8}\right).
\end{equation}
\noindent Therefore the total drag is
\begin{align}
	F_d &=F_d^r+F_d^a=\frac{3\pi\mu l_n RU}{\theta}\left[1-\epsilon_2\left(\frac{3}{\pi^2}-
	\frac{1}{4}\right)\right]\notag\\
	&\approx\frac{3\pi\mu l_n RU}{\theta}\left(1-\frac{\epsilon_2}{20}\right).
\end{align}

Next we investigate the resultant driving force $F$ when 
droplet deformation is involved in. 
The resultant force at the receding contact line reads 
\begin{align}
	F^r &=  2\int_0^{\pi/2}[\gamma_0+\mathit{\Gamma} r(\varphi)\cos\varphi]\cos\varphi Rd\varphi\notag\\
	&=\pi R^2\mathit{\Gamma}/2+2\gamma_0 R, 
\end{align}
\noindent and the resultant force at the advancing contact line is 
\begin{align}
	F^a=& 2\int_{\pi/2}^{\pi}\gamma n_x ds \notag\\
	=&  2\int_{\pi/2}^{\pi}[\gamma_0+\mathit{\Gamma} r(\varphi)\cos\varphi] n_x ds \notag\\
	\approx& 2R\int_{\pi/2}^{\pi}\left(\gamma_0\cos\varphi+\mathit{\Gamma} R\cos^2\varphi\right)d\varphi\notag\\
	&+2R\int_{\pi/2}^{\pi}\epsilon_2[\gamma_0f(\varphi)\cos\varphi+\gamma_0f'(\varphi)\sin\varphi]d\varphi\notag\\
	&+2R\int_{\pi/2}^{\pi}\epsilon_2[2\mathit{\Gamma} R\cos^2\varphi f(\varphi)\notag\\
	&+\mathit{\Gamma} R\cos\varphi\sin\varphi f'(\varphi)]d\varphi\notag\\
	=& -2\gamma_0 R+\pi R^2(1+\epsilon_2)\mathit{\Gamma}/2. 
\end{align}
\noindent Hence the resultant driving force reads
\begin{equation}
	\begin{aligned}
		F= F^r+ F^a=\pi R^2\mathit{\Gamma}(1+\epsilon_2/2).
	\end{aligned}
\end{equation}

The surface tension gradient of the droplet is 
\begin{align}
	\mathit{\Gamma}&\sim\frac{1}{(L+2R+R\epsilon_2)^2}\notag\\
	&=\frac{1}{(L+2R)^2}\frac{1}{[1+R\epsilon_2/(L+2R)]^2}\notag\\
	&=\frac{1}{(L+2R)^2} \left[1-\frac{2R\epsilon_2}{L+2R}+o(\epsilon_2)\right],
\end{align}
\noindent and we can get $\mathit{\Gamma}=\mathit{\Gamma}_0(1-\epsilon_2)$ and total resultant force 
\begin{equation}
	F=\pi R^2\mathit{\Gamma}_0(1+\epsilon_2/2)(1-\epsilon_2)\approx\pi R^2\mathit{\Gamma}_0(1-\epsilon_2/2),
\end{equation}
where $\mathit{\Gamma}_0$ denotes surface tension gradient 
for the droplet without deformation. 
\noindent Comparing two cases with and without droplet deformation, 
\begin{align}
	\pi R^2\mathit{\Gamma}_0=&\frac{3\pi\mu l_n RU_0}{\theta},\notag\\
	\pi R^2\mathit{\Gamma}_0(1-\frac{\epsilon_2}{2})=&\frac{3\pi\mu l_n RU}{\theta}\left(1-\frac{\epsilon_2}{20}\right), 
\end{align}
we get $U / U_0 = (1-\epsilon_2/2) / (1-\epsilon_2/20)\approx 1-9\epsilon_2/20$, 
where $U_0$ denotes velocity of the droplet without deformation.
Finally, the change in velocity due to small deformation of 
droplet is $|(U-U_0) / U_0| < 5\%$, 
which is negligible. 

\section{Driven force between evaporating droplets with different radius ratios.}
\begin{figure}[htbp]
	\centering
	\includegraphics[width=\linewidth]{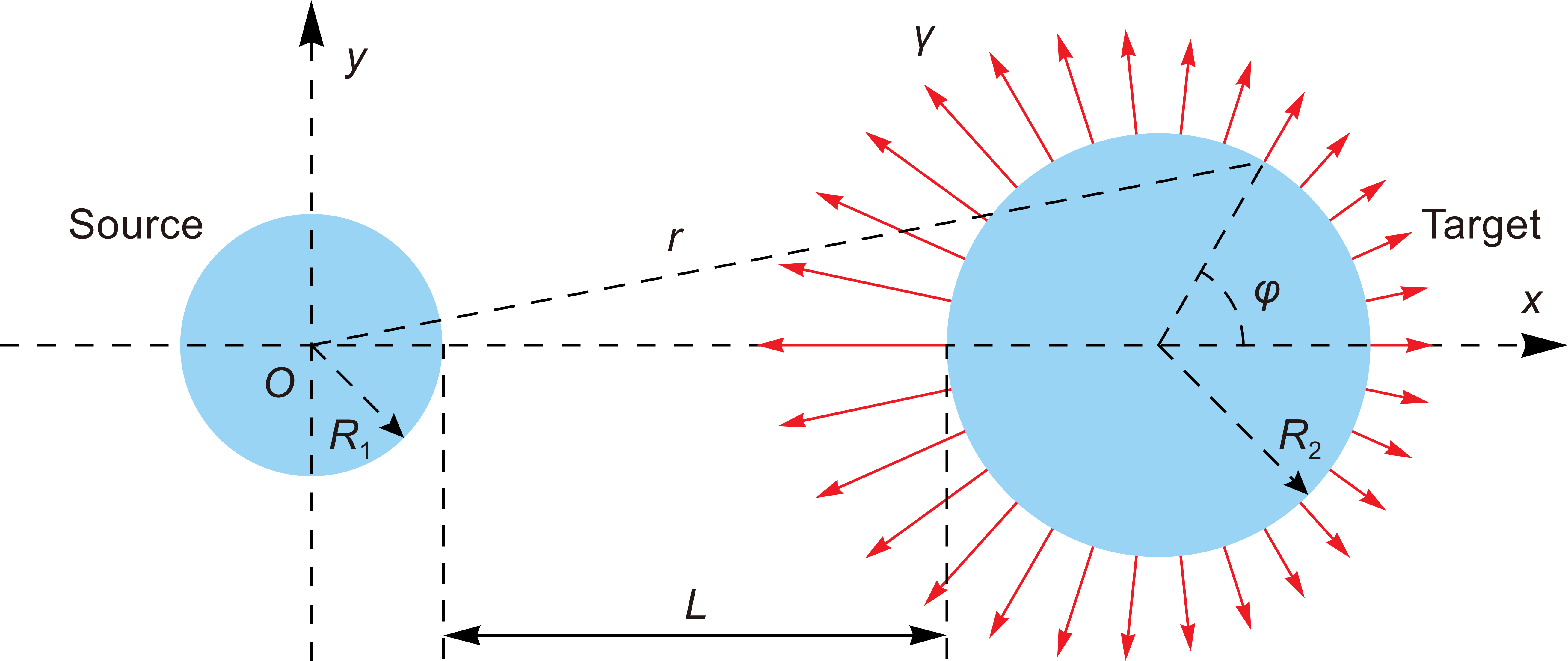}
	\caption{\label{fig:s9}
		Attraction between two unequal size binary droplets.}
	\label{diffR}
\end{figure}

When the ``source" and ``target" droplets are different in size (Fig. S\ref{diffR}),
the resultant driving force on ``target" droplet reads
\begin{align}
	F=&\int_{0}^{2\pi}\gamma\cos{\varphi} R_2d\varphi\notag\\
	=&\int_{0}^{2\pi}\gamma_d\cos\theta\cos\varphi R_2d\varphi\notag\\
	=&\int_{0}^{2\pi}\gamma_d(m\phi+b)\cos\varphi R_2d\varphi\notag\\
	=&\gamma_d m_p \phi_{\infty} R_2\int_{0}^{2\pi}\phi\cos\varphi d\varphi, 
\end{align}
\noindent substituting $\phi=\lambda(1-\phi_\infty)R_1/r+\phi_\infty$, 
we get 

\begin{align}
	F= &\gamma_d m_p \phi_{\infty} R_2\int_{0}^{2\pi}\Big[\frac{\lambda(1-\phi_\infty)R_1}{r}+\phi_\infty\Big]\cos\varphi d\varphi \notag\\
	= &\gamma_d  m_p \phi_{\infty} \lambda R_1R_2(1-\phi_\infty)\int_{0}^{2\pi}\frac{\cos\varphi}{r}d\varphi.
\end{align}
\noindent We introduce $\epsilon_3 = R_2/(L+R_1+R_2)$ to simplify $1/r$ 
\begin{align}
	\frac{1}{r} =&\frac{1}{\sqrt{(L+R_1+R_2+R_2\cos\varphi)^2+(R_2\sin\varphi)^2}}\notag \\
	= &\frac{1}{L+R_1+R_2}\frac{1}{\sqrt{1+2\epsilon_3\cos{\varphi}+\epsilon_3^2}}\notag \\
	=&\frac{1}{L+R_1+R_2}\sum_{n=0}^{\infty}{P_n(-\cos\varphi)\epsilon_3^n}\notag\\
	= & \frac{1}{L+R_1+R_2}\Big[1-\epsilon_3\cos\varphi+\frac{\epsilon_3^2}{2}(3\cos^2\varphi-1)\notag\\
	&+  \frac{\epsilon_3^3}{2}(-5\cos^3\varphi+3\cos\varphi)+o(\epsilon_3^3)\Big]\notag\\
	\approx & \frac{1}{L+R_1+R_2}\Big[1-\epsilon_3\cos\varphi+\frac{\epsilon_3^2}{2}(3\cos^2\varphi-1)\notag\\
	&+  \frac{\epsilon_3^3}{2}(-5\cos^3\varphi+3\cos\varphi)\Big]. 
\end{align}

\begin{figure}[htbp]
	\centering
	\includegraphics[width=0.8\linewidth]{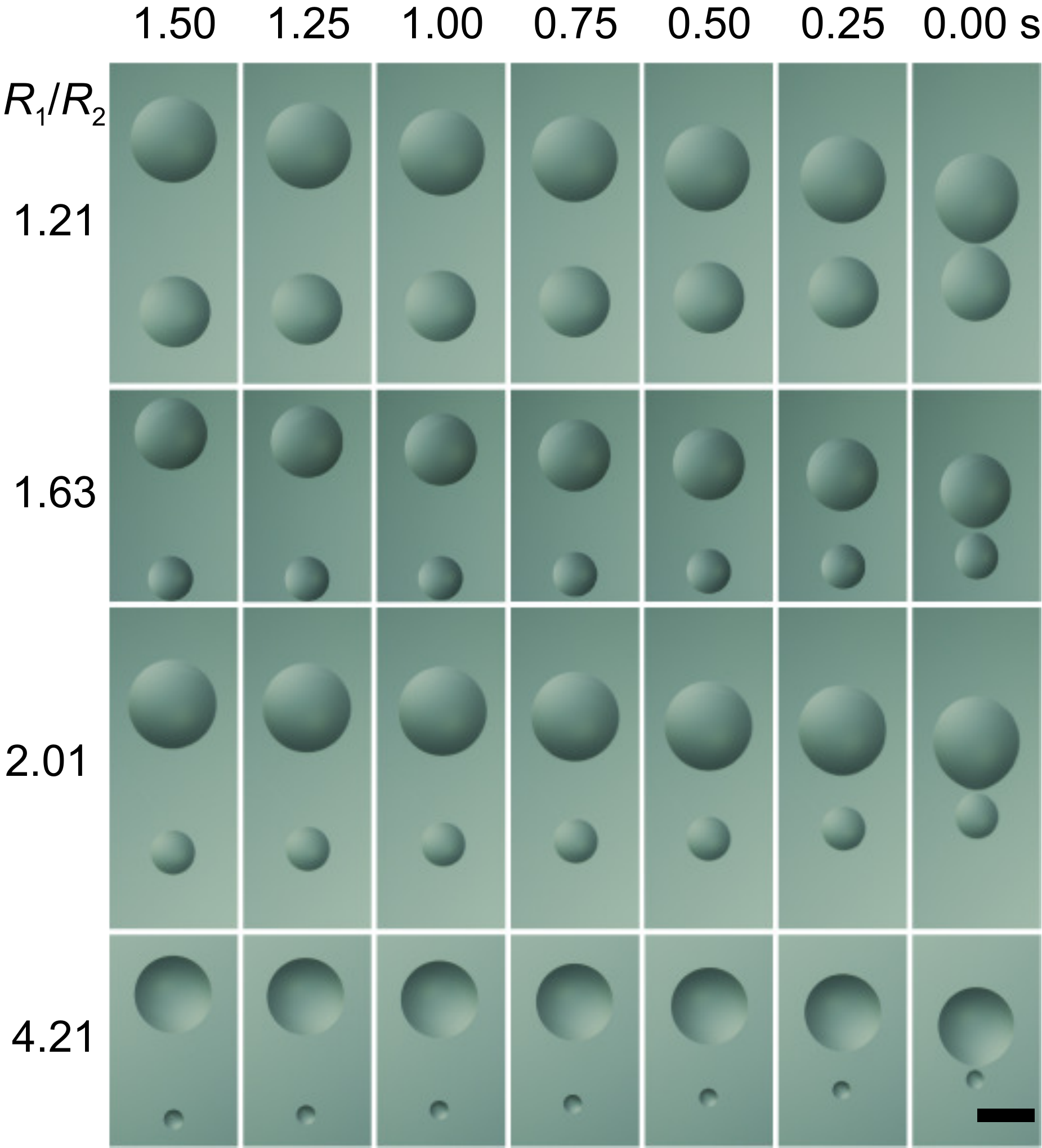}
	\caption{Top view showing attraction of two 10 wt\% PG aqueous sessile droplets with different spreading radius ratio $\zeta = R_1/R_2$. 
		The time shown is for $\tau-t$. 
		Sequence of images spaced by 0.25 s. 
		The scale bar is 2 mm.}
	\label{fig:s10}
\end{figure}
\begin{figure}[htbp]
	\centering
	\includegraphics[width=0.8\linewidth]{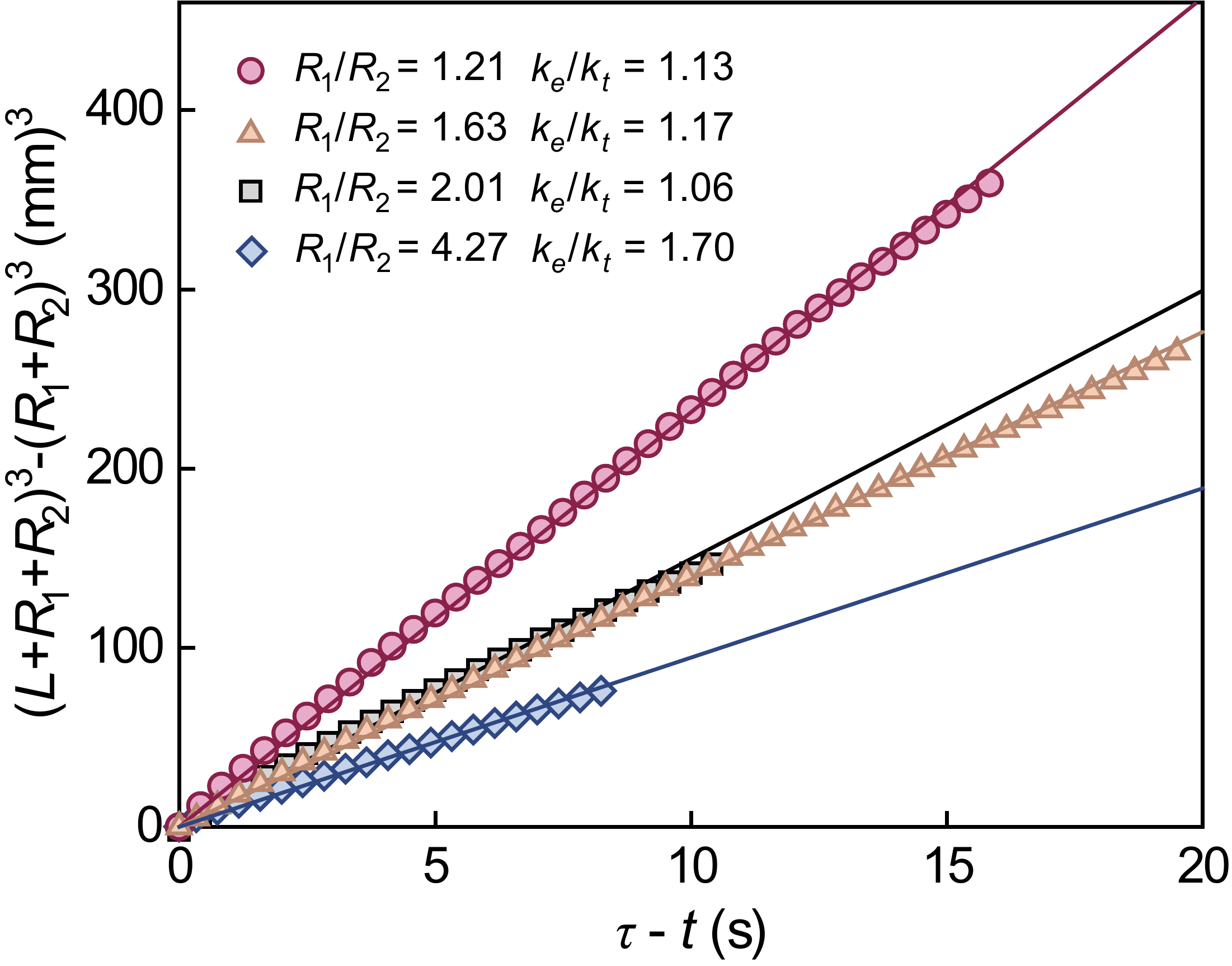}
	\caption{\label{fig:s11}
		Data fall into straight lines implying an inverse-square force between 10 wt\% PG droplets of different sizes.}
	\label{4lines}
\end{figure}

Hence the integral item $\int_{0}^{2\pi}\cos\varphi /rd\varphi$ reads
\begin{align}
	\int_{0}^{2\pi}&\frac{\cos\varphi}{r}d\varphi\notag\\
	\approx&\frac{1}{L+R_1+R_2}\int_{0}^{2\pi}\cos\varphi\Big[1-\epsilon_3\cos\varphi \notag\\
	&+\frac{\epsilon_3^2}{2}(3\cos^2\varphi-1)+\frac{\epsilon_3^3}{2}(-5\cos^3\varphi+3\cos\varphi)\Big]d\varphi\notag\\
	=&-\frac{1}{L+R_1+R_2}(\pi\epsilon_3+{3\pi\epsilon_3^3}/{8})\notag\\
	=&-\frac{\pi R_2}{(L+R_1+R_2)^2}-\frac{3\pi R_2^3}{8(L+R_1+R_2)^4}.\label{eqn:43}
\end{align}
Ignoring higher-order terms, the resultant force reads 
\begin{equation}
	F=-\frac{\gamma_d m_p \phi_{\infty} \lambda \pi R_1R_2^2(1-\phi_\infty)}{(L+R_1+R_2)^2}.
	\label{f_IgnoreHigh}
\end{equation}

The deviation of Eq. \ref{f_IgnoreHigh} from Eq. \ref{eqn:43} is $3\epsilon_3^2/8$. 
Denoting the spreading radius ratio of ``source" droplet 
and ``target" droplet as $\zeta = R_1/R_2$, 
the definition of $\epsilon_3$ gives $\epsilon_3 < 1/(1 + \zeta)$. 
From the perspective that the ``target'' droplet behaves simultaneously as a ``source'' droplet, 
we have another spreading radius ratio which is $1/\zeta$ 
and the corresponding relationship $\epsilon_3 < 1/(1+1/\zeta)$. 
Let us say the interference of high-order terms in Eq. \ref{eqn:43} can be ignored if the deviation satisfies
\begin{align}
	\frac{3\epsilon_3^2}{8}&<\frac{3}{8}\frac{1}{(1+\zeta)^2}<20\%\ \mathrm{and} \notag\\
	\frac{3\epsilon_3^2}{8}&<\frac{3}{8}\frac{1}{(1+1/\zeta)^2}<20\%. 
\end{align}
\noindent Correspondingly, the condition for Eq. \ref{f_IgnoreHigh} to accurately predict the driving force 
is $0.37\leqslant R_1/R_2\leqslant 2.70$. 
To verify above analysis, 
experiments were conducted for droplets with different sizes and part of the results were shown in Fig. S\ref{fig:s10} and Fig. S\ref{4lines}. 
When radius ratios $\zeta < $ 3, 
the calculated $k_e / k_t$ is nearly equal to 1. 

\section{Inverse-square force between floating binary droplets on a liquid free surface.}
\begin{figure}[tbp]
	\centering
	\includegraphics[width=\linewidth]{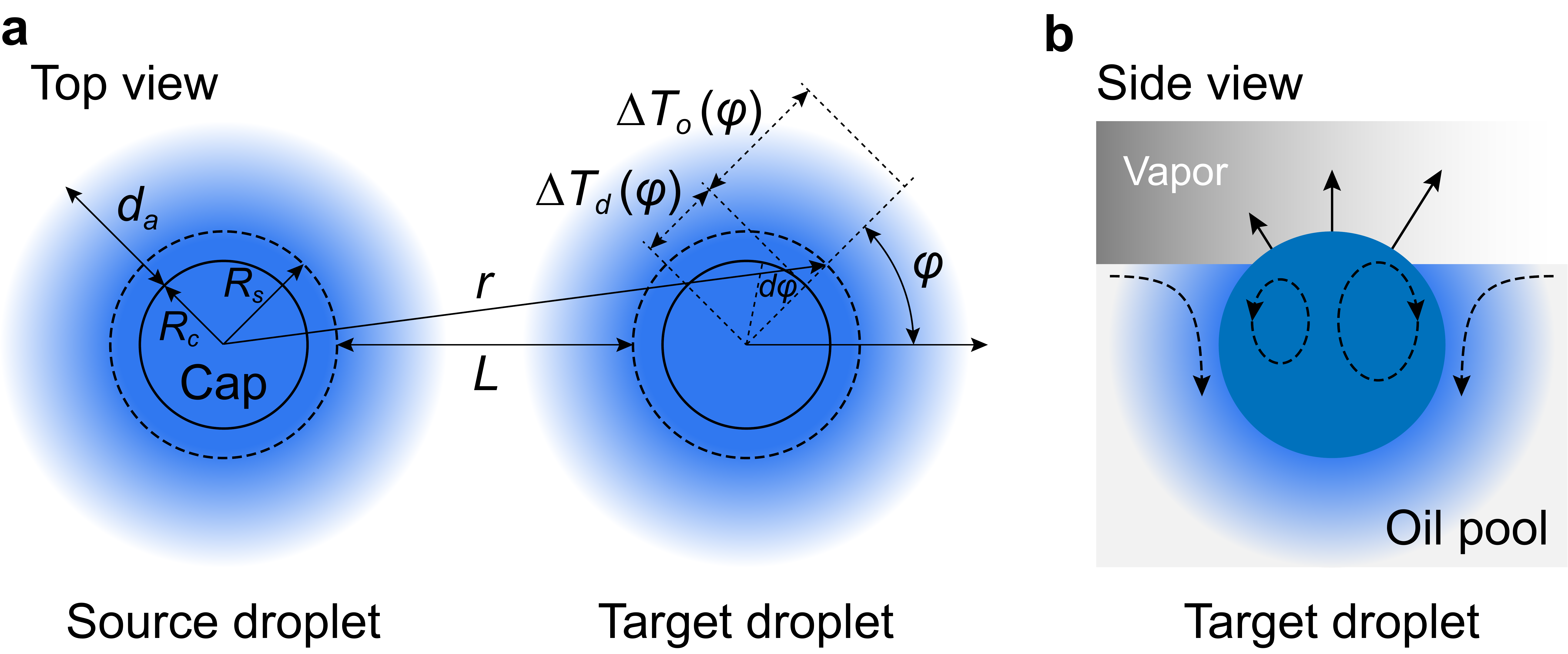}
	\caption{\label{tuan}
		A schematic diagram showing the geometric parameters and temperature field in the floating droplet system. 
		A copy sketch to fig. 4 in Liu \& Tran \cite{liu2018}. 
		(a) Top-view schematic of the system consisting of a ``source'' and ``target'' droplet partially immersed into a liquid pool. $R_s$ and $R_c$ are droplet radius and cap radius, respectively. 
		$d_a$ denotes a length scale within which the oil 
		around the droplet will be thermally affected by the 
		lower temperature of the cap.  
		(b) Side-view schematic of the target droplet. 
		Dashed arrows indicate convective flows. 
		Solid arrows indicate nonuniform evaporation across the cap of the target droplet.}
\end{figure}

Two IPA aqueous droplets partially submersed in an immiscible silicone oil pool will attract each other \cite{liu2018}. 
Nonuniform vapor density produces a temperature gradient over the cap of the floating droplet and subsequent a thermocapillary stress contributing to droplet motion {\color{black}(Fig. S\ref{tuan}, 
	a copy sketch to fig. 4 in Liu \& Tran \cite{liu2018})}. 
The evaporation rate per unit area is estimated as 
\begin{equation}
	\Dot{m}_m\left(\varphi\right)\sim1-\frac{R_c}{r}, 
\end{equation}
\begin{equation}
	r=\sqrt{\left(L+2R_s+R_c\cos\varphi\right)^2+\left(R_c\cos\varphi\right)^2}, 
\end{equation}
\noindent where $R_c$ is the radius of droplet cap, 
$r$ is the distance from the center of ``source" droplet, 
and $R_s$ is the droplet radius. 

The temperature difference 
$\Delta T_d\left(\varphi\right)\sim\left[\Dot{m}_m\left(\varphi\right)\right]^{1/2}$ 
across the droplet cap generates a thermocapillary stress $\tau_d\left(\varphi\right)\approx\left(\gamma_T\Delta T_d\left(\varphi\right)\right)/R_c\sim\Delta T_d\left(\varphi\right)$, 
where $\gamma_T=\partial\gamma/\partial T$ is derivative of 
the surface tension $\gamma$ with respect to $T$.  
Meanwhile, the temperature difference 
$\Delta T_o\left(\varphi\right)\sim\left[\Dot{m}_m\left(\varphi\right)\right]^{1/2}$ 
across the surrounding thermally affected area with 
characteristic distance $d_a$ generates the stress $\tau_o\left(\varphi\right)\approx\left(\gamma_T\Delta T_o\left(\varphi\right)\right)/d_a\sim\Delta T_d\left(\varphi\right)$. 
Hence the thermocapillary stress gradient of droplet cap $\mathit{\Gamma}_d$ and surrounding thermally affected area $\mathit{\Gamma}_o$ are 
\begin{align}
	&\mathit{\Gamma}_{d}=\frac{d \tau_{d}}{d r} \sim \frac{d}{d r}\left(1-\frac{R_{c}}{r}\right)^{1 / 2} \sim \frac{1}{\left(1-R_{c} / r\right)^{1 / 2}} \frac{1}{r^{2}},\notag \\
	&\mathit{\Gamma}_{o}=\frac{d \tau_{o}}{d r} \sim \frac{d}{d r}\left(1-\frac{R_{c}}{r}\right)^{1 / 2} \sim \frac{1}{\left(1-R_{c} / r\right)^{1 / 2}} \frac{1}{r^{2}},
\end{align}
respectively. 
When the distance $r\gg R_c$, 
the thermocapillary stress of droplet cap and surrounding 
thermally affected area can be expressed as 
\begin{align}
	F_{d} &\sim \mathit{\Gamma}_{d} R \cdot \pi R^{2} \sim \mathit{\Gamma}_{d} \sim \frac{1}{r^{2}},\notag \\
	F_{o} &\sim \mathit{\Gamma}_{o} R \cdot \pi R^{2} \sim \mathit{\Gamma}_{o} \sim \frac{1}{r^{2}}. 
\end{align}
\noindent The resultant thermocapillary stress is
\begin{equation}
	F=F_d+F_o\sim\frac{1}{r^2}. 
\end{equation}

Eventually, we obtain an inverse-square attractive force 
between floating binary droplets on a liquid free surface. 
The viscous drag is $F_d \sim U$ and the kinematic 
relationship for droplet motion can be described as 
\begin{equation}
	\left(L_0+2R\right)^3-\left(L+2R\right)^3\sim t, 
	\label{TranKR}
\end{equation}
\noindent where $L_0$ is the initial distance between droplets. 
Experimental verification of Eq. \ref{TranKR} is shown in 
Fig. 4d in the main text. 
When the distance between droplets is smaller than 
a critical distance, 
the thermally affected regions of the droplet couples overlap 
and brings other force into play and interferes droplet motion. 

\section{Comparison with the law of universal gravitation.}
\begin{figure}[htbp]
	\centering
	\includegraphics[width=0.66\linewidth]{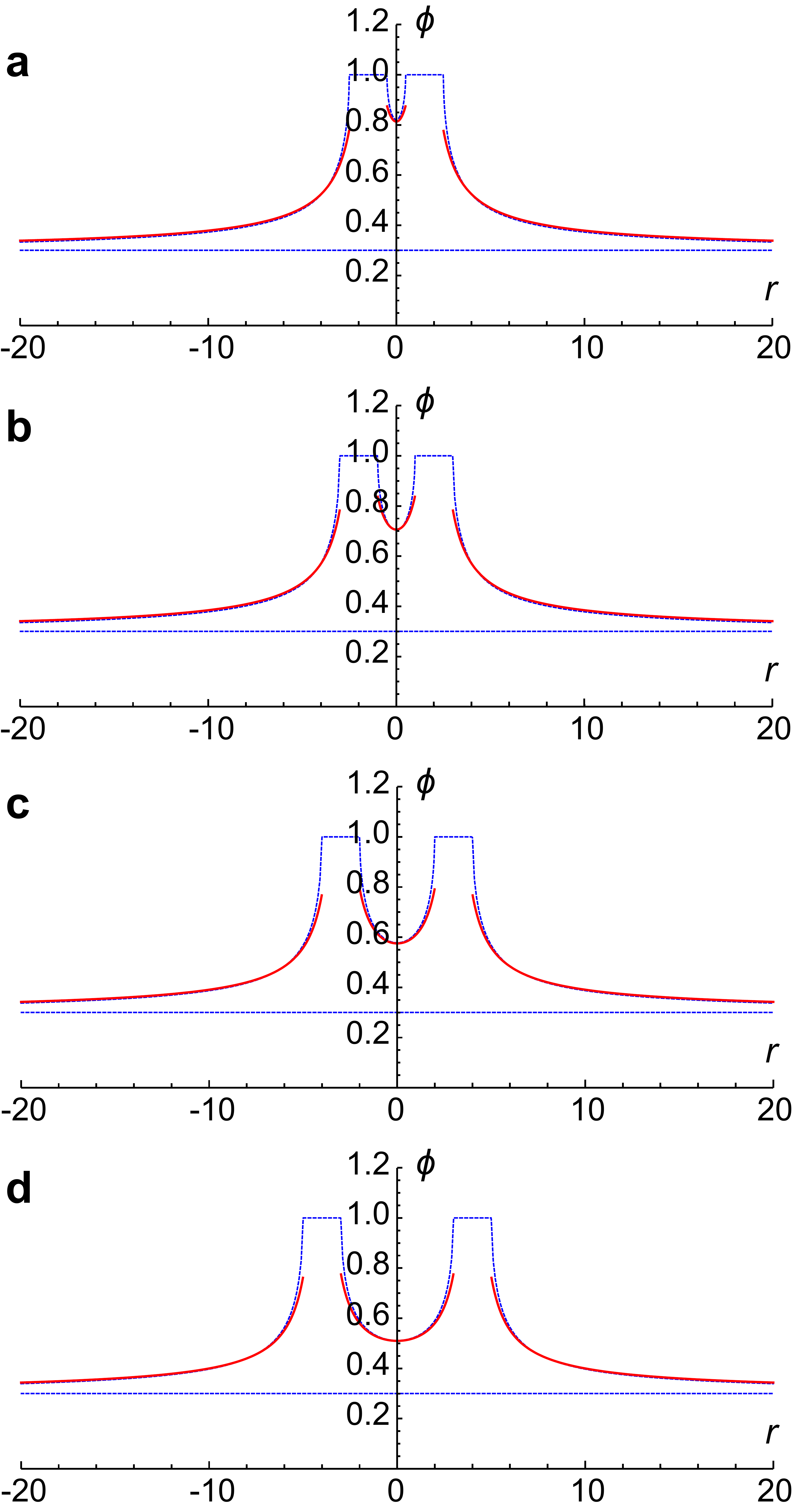}
	\caption{\label{fig:s13}
		Verification for superposition principle. 
		(a), (b), (c) and (d) compares the numerical 
		results (blue dash lines) with theoretical results 
		(red solid lines), for total humidity 
		distribution from two ``source'' droplets  
		$L=R$, $2R$, $4R$ and $6R$ apart, respectively.}
\end{figure}
The gravitational potential of a ``source" with mass $M$, 
and the gravitation of a ``target" with mass $m$ in the 
gravitational field of above ``source" are 
\begin{align}
	E=\int_{\infty}^{r} \frac{G M}{r^{2}} d r&=-\left.\frac{G M}{r}\right|_{\infty} ^{r}=-\frac{G M}{r}, \notag\\
	F_m&=\frac{GMm}{r^2},
	\label{eqn:121}
\end{align}
\noindent respectively, 
where $G$ is the gravitational constant. 
The gravitation can be written as mass multiplied by the 
gradient of the gravitational potential $F_m=-m\nabla E$. 
Here, we define a humidity potential induced by evaporation 
$E^{\prime} = -G^{\prime} M^{\prime} / r = -\gamma_d m_p\phi_\infty \lambda(1 - \phi_\infty)R_1 / r$, 
where $G^{\prime} = d\gamma / d\phi = \gamma_d m_p \phi_\infty$
is a constant denoting the derivative of surface tension 
to environment humidity, 
$M^{\prime} = \lambda(1 - \phi_\infty) R_1$ represents 
the decay coefficient of humidity away from the ``source" droplet, 
$m^{\prime} = \pi R_2^2$ represents the wetting area of 
the ``target" droplet, respectively. 
In an identical coordinate system as the universal gravitation, 
the resultant force exerted on the ``target" droplet 
in the humidity field of the ``source" droplet reads 
\begin{align}
	F&=\pi R_2^{2} \mathit{\Gamma}=\pi R_2^{2} \nabla \gamma=\pi R_2^{2} \frac{d \gamma}{d \phi} \nabla \phi\notag\\
	&=\gamma_{d} m_p \phi_{\infty} \pi R_2^{2} \frac{\lambda\left(1-\phi_{\infty}\right) R_1}{r^{2}}. 
	\label{eqn:122}
\end{align}
\noindent

Next we investigate whether $F$ satisfies 
the superposition principle. 
For an ideal small ``target" object with mass $m$ in the 
vicinity of two ``source" objects with mass $M$, 
and assume the existence of $m$ does not affect the 
gravitational potential fields of the ``source'' objects, 
the resultant gravitational potential and gravitation read 
\begin{align}
	E&=\int_{\infty}^{\vec{r}}\left(\frac{G M}{r_{1}^{2}} \overrightarrow{n_{1}}+\frac{G M}{r_{2}^{2}} \overrightarrow{n_{2}}\right) \cdot d \vec{r}\notag\\
	&=-\frac{G M}{r_{1}}-\frac{G M}{r_{2}},\\ 
	F_m&=\frac{G M m}{r_{1}}\overrightarrow{n_{1}}+\frac{G M m}{r_{2}}\overrightarrow{n_{2}},
\end{align}
\noindent respectively, 
where $r_1$ and $r_2$ are distances between ``target'' and the two ``source'' objects, $\vec{n}_1$ and $\vec{n}_2$ are the unit vectors pointing from center of ``target'' to the center of two ``source'' objects, respectively. 

The superposition of the humidity field is not that 
straightforward as there is an upper limit for humidity, 
which is 1. 
We invoke a coefficient $\lambda'$ to guarantee the 
validity of humidity range. 
Figure S\ref{fig:s13} shows the numerical results of the 
humidity distribution on the central axis of two ``source'' 
droplets with different spacing, 
verifying the feasibility of describing humidity distribution as 
\begin{equation}
	\phi=\frac{\lambda'\left(1-\phi_{\infty}\right) R}{r_{1}}+\frac{\lambda'\left(1-\phi_{\infty}\right) R}{r_{2}}+\phi_{\infty},
	\label{eqn:124}
\end{equation}
\noindent where $\lambda' = 0.55,\ 0.58,\ 0.59,\ 0.60$ for $L=R$, $2R$, $4R$ and $6R$, respectively. 
Analogously, the resultant force exerted on a ``target" droplet with spreading radius $a$ in the humidity field of two ``source"  
droplets with spreading radius $R$ can be expressed in a 
similar form to the law of universal gravitation (Fig. 5 in the main text) 
\begin{align}
	F =&\pi a^{2} \mathit{\Gamma}=\pi a^{2} \nabla \gamma =\pi a^{2} \frac{d \gamma}{d \phi} \nabla \phi\notag\\
	=&\pi a^{2} \frac{d \gamma}{d \phi} \nabla\left(\phi_{1}+\phi_{2}\right) \notag\\
	=&\gamma_{d} m_p \phi_{\infty} \pi a^{2} \frac{\lambda'\left(1-\phi_{\infty}\right) R}{r_{1}^{2}} \overrightarrow{n_{1}}\notag\\
	&+\gamma_{d} m_p \phi_{\infty} \pi a^{2} \frac{\lambda'\left(1-\phi_{\infty}\right) R}{r_{2}^{2}} \overrightarrow{n_{2}} \notag\\
	=&\frac{G^{\prime} M^{\prime} m^{\prime}}{r_{1}^{2}} \overrightarrow{n_{1}}+\frac{G^{\prime} M^{\prime} m^{\prime}}{r_{2}^{2}} \overrightarrow{n_{2}},
\end{align}
where $G^{\prime}=\gamma_d m_p \phi_\infty$, $M^{\prime}=\lambda'(1-\phi_\infty)R$, 
$m^{\prime}=\pi a^2$, 
$\vec{n}_1$ and $\vec{n}_2$ are the unit vectors pointing from 
center of ``target'' to the center of two ``source'' droplets, 
respectively. 
Clearly, Eq. S67 indicates that the driving force in this study
satisfies the superposition principle. 

\end{document}